\documentclass[leqno,12pt,a4paper]{article}
\usepackage{setspace}
\usepackage{latexsym}
\usepackage{amsmath}
\usepackage{amssymb}
\usepackage{mathtools}
\usepackage{bbm}
\usepackage{amsthm}
\usepackage{marvosym}

\usepackage[]{algorithm,algpseudocode}
\algnewcommand\algorithmicforeach{\textbf{for each}}
\algdef{S}[FOR]{ForEach}[1]{\algorithmicforeach\ #1\ \algorithmicdo}

\usepackage{graphicx}
\usepackage{epsfig}
\usepackage{subcaption}
\usepackage{tabularx}
\newcolumntype{L}[1]{>{\raggedright\arraybackslash}p{#1}} % linksbÃ¼ndig mit Breitenangabe
\newcolumntype{C}[1]{>{\centering\arraybackslash}p{#1}} % zentriert mit Breitenangabe
\newcolumntype{R}[1]{>{\raggedleft\arraybackslash}p{#1}} % rechtsbÃ¼ndig mit Breitenangabe

\usepackage{array}

\allowdisplaybreaks
\usepackage{tikz}
\usetikzlibrary{shapes,arrows,patterns,backgrounds,fit,arrows.meta}
\usepackage{pgfplots}
\pgfplotsset{
  compat=newest,
  width=0.65\textwidth,
  height=0.4\textwidth,
  tick align=outside,
  tick pos=left,
  scaled ticks=false,
  grid=major,
  grid style={dotted},
}
\pgfkeys{/pgf/number format/.cd,
  1000 sep={\,}, % 1000 separation with distance instead of comma
  min exponent for 1000 sep=4,
}

\usepackage{siunitx}
\usepackage{makecell}

\usepackage{comment}

\newcommand{\boldface}[1]{\boldsymbol{#1}}

\newcommand{\bfb}{\boldface{b}}

\newcommand{\bfy}{\boldface{y}}

\newcommand{\bfC}{\boldface{C}}

\newcommand{\bfF}{\boldface{F}}

\newcommand{\bfW}{\boldface{W}}

\newcommand{\calI}{\mathcal{I}}

\newcommand{\calL}{\mathcal{L}}

\newcommand{\calO}{\mathcal{O}}

\newcommand{\calR}{\mathcal{R}}

\usepackage{bbold}

\newcommand{\dsR}{\mathbb{R}}

\newcommand{\be}{\begin{equation}}
\newcommand{\ee}{\end{equation}}
\newcommand{\bea}{\begin{eqnarray}}
\newcommand{\eea}{\end{eqnarray}}
\newcommand{\bes}{\begin{equation*}}
\newcommand{\ees}{\end{equation*}}
\newcommand{\beas}{\begin{eqnarray*}}
\newcommand{\eeas}{\end{eqnarray*}}

\newcommand{\tr}[1]{\mathrm{tr} \, #1}

\newcommand{\trC}{\tr[\bfC]}

\newcommand{\norm}[1]{\left\lVert#1\right\rVert}

\usepackage{lipsum}
\usepackage{placeins}

\definecolor{RUBblue}{rgb}{0.0470588,0.262745,0.411765}
\definecolor{lightgray}{gray}{0.92}
\definecolor{blau}{rgb}{0,0.25,1.0}
\definecolor{RUBCDblue}{rgb}{0.062745,0.305882,0.545098}
\definecolor{RUBCDgreen}{rgb}{0.517647,0.741176,0.0}
\definecolor{RUBCDgray}{rgb}{0.815686,0.815686,0.807843}
\definecolor{RUBCDgrayDark}{rgb}{0.615686,0.615686,0.607843}

\makeatletter

\def\env@cases#1{%
  \let\@ifnextchar\new@ifnextchar
  \left\lbrace\def\arraystretch{1.2}%
  \array{@{}#1@{\quad}l@{}}}
\makeatother

\usepackage{booktabs}

\RequirePackage{relsize}
\newfont{\Sf}{cmssbx10 scaled 2074}

\newcommand{\figref}[1]{Fig.~\ref{#1}}
\newcommand{\tabref}[1]{Tab.~\ref{#1}}

\newcommand{\rev}[1]{\textcolor{black}{#1}}
\newcommand{\revv}[1]{\textcolor{black}{#1}}

\usepackage[
  colorinlistoftodos,
  backgroundcolor=green,
  linecolor=green,
  textsize=footnotesize
]{todonotes}

\usepackage[backend=bibtex,firstinits=true, url=false]{biblatex}
\usepackage{import}
\graphicspath{{plots/EUCLID}, {plots/network_architecture_latex_bundle}}

\begin{document}
\title{Sparse Constitutive Neural Networks for Hyperelasticity
}
\date{} 
\maketitle
{\large
\noindent{Hendrik}  Geisler$^{1}$, Clément Jailin$^{2}$, Emmanuel Baranger$^{2}$ \\[0.5mm]
}
1: Leibniz University Hannover, Institute of Continuum Mechanics, Hannover, Germany\\[2mm]
2: Université Paris-Saclay, CentraleSupélec, ENS Paris-Saclay, CNRS, LMPS - Laboratoire de Mécanique Paris-Saclay, Gif-sur-Yvette, France\\[2mm]
\Letter \hspace{0.1cm} geisler@ikm.uni-hannover.de

\section*{Abstract}
Constitutive model identification aims to recover compact, interpretable models that accurately describe material behavior. In this work, we identify hyperelastic constitutive models using artificial neural networks with sparsity-promoting constraints and regularization. The proposed method combines the expressiveness of neural networks with the structural interpretability of sparse constitutive models, such as those identified by EUCLID, by representing the strain energy \revv{density} as a sum of products of univariate functions of the invariants.
The resulting decomposition exposes the contributions of individual invariants and their interactions.
The method is demonstrated on synthetic data in supervised and unsupervised learning scenarios, showing that it can accurately identify interpretable constitutive models with few active modes.

\section{Introduction}
Data-driven discovery enables hyperelastic constitutive models to be learned directly from measurements, with limited human intervention and without prescribing a functional form a priori.
\rev{Especially when the material response is complex or the available data are noisy or incomplete, handcrafting a constitutive model can be challenging. In such cases, data-driven approaches can provide a more flexible and efficient alternative to traditional methods.}

Different approaches for data-driven constitutive modeling have been proposed in the literature.
An overview \rev{for fully data-driven approaches} can be found in \cite{romer_reduced_2025}.
These approaches can be broadly grouped into two categories: sparse identification and neural networks. Sparse identification methods aim to discover governing equations by combining a library of candidate functions with sparse regression techniques. The Sparse Identification of Nonlinear Dynamical Systems (SINDy) framework is a prominent example, demonstrating that governing equations can often be recovered from data \cite{brunton_discovering_2016}. In the context of constitutive modeling, EUCLID \cite{flaschel_unsupervised_2021} extends the SINDy approach. Symbolic regression has also been used to discover interpretable constitutive models \cite{kissas_language_2024, abdusalamov_automatic_2023, bahmani_physicsconstrained_2024}.

While these approaches can yield concise and interpretable equations, their success heavily depends on the choice of the function library and the discrete model-selection procedure. If the relevant functional form is absent from the library, it cannot be recovered, and small changes in noisy data can lead to discontinuous changes in the identified model\revv{, i.e., a jump between mathematical expressions}.

An alternative approach is the use of neural networks. Physics-augmented neural networks
can approximate highly nonlinear responses without enumerating a library of
candidate functions, and energy-based architectures can incorporate physical
constraints such as objectivity and thermodynamic consistency
\cite{ghaboussi_knowledge_1991,linka_thermodynamics-based_2021,linden_neural_2023}. The learning procedures of these networks for learning from experimental data are investigated in \cite{thakolkaran_nn-euclid_2022, jailin_experimental_2024, benady_nnmcre_2024}.

Several architectures have been proposed to incorporate these physical constraints directly into the neural network. Input-convex neural networks (ICNNs) impose convexity with respect to selected inputs through architectural constraints on the network weights \cite{amos_input_2017}. In hyperelasticity, this idea has been used to construct energy potentials that satisfy objectivity, material symmetry, polyconvexity, and thermodynamic consistency by design \cite{linden_neural_2023}. Constitutive artificial neural networks (CANNs) \cite{linka_constitutive_2021, linka_new_2023, alheit_cann-euclid_2026} provide a related energy-based framework in which the strain energy is represented by a structured combination of invariant-dependent functions. However, the higher degree of interpretability is achieved at the cost of reduced expressiveness, as the functional form is restricted to a specific combination of invariant-dependent functions.
The relative advantages and limitations of these architectures have been examined in comparative studies \cite{tac_benchmarking_2024}. 

The flexibility of these models is useful for interpolation between training data points, which is a key requirement when the constitutive law is evaluated inside a finite-element simulation. However, even physically constrained neural networks may contain many coupled contributions whose parameters do not directly reveal which invariants interact or which mechanisms dominate the strain energy. A representation that combines the expressive capacity of neural networks with an explicitly sparse decomposition is therefore needed.

Different approaches have been proposed to increase the interpretability of neural networks. 
Kolmogorov-Arnold-Networks (KANs) learn univariate functions that can be visualized more easily \cite{thakolkaran_can_2025}. However, multilayer compositions can obscure the underlying relationships.
The network structure can be sparsified during training \cite{fuhg_extreme_2024}. Nevertheless, sparsity of network parameters does not necessarily produce a transparent decomposition of the strain energy. 
A large variety of work has been done on interpretable machine learning, see for example \cite{rudin_interpretable_2021}. These techniques did not yet find widespread application in constitutive modeling.

These approaches highlight a tension between expressiveness and interpretability.
Interpretability is a broad concept with different definitions. Here, we adopt the notion of transparency discussed by Lipton \cite{lipton_mythos_2018}. This notion includes simulatability, or the ability to understand the model as a whole, and decomposability, or the ability to understand its individual components. A sparse decomposition in which each term is associated with a small number of invariant coordinates can therefore provide more useful scientific information than a black-box predictor with comparable error. \rev{Although such decompositions are generally not unique, they can provide an interpretable representation of the learned response,} support comparison with established material models, and identify invariant interactions that are sufficient within the chosen representation.

In this work, we combine the approximation capability of neural networks with
the structural interpretability of sparse constitutive models. We represent the
strain energy as a sum of products of univariate functions of the invariants \revv{similar to decompositions in POD and PGD \cite{chatterjee_introduction_2000, ladeveze_latin_2010}}.
The functions are learned by one-dimensional input-convex subnetworks, while
non-negativity constraints, regularization, and mode pruning promote a sparse
decomposition. 

The resulting model has three intended properties. First, it retains the
expressiveness of neural networks and does not require a complete library of
candidate functions. Second, its active modes and one-dimensional components
can be visualized and related to familiar constitutive terms. Third, its
energy-based construction preserves a physically meaningful strain-energy
representation and supports interpolation throughout the deformation domain.
We assess these properties in supervised identification experiments with known
synthetic energies and in an unsupervised setting based on synthetic
full-field deformation and reaction-force data.

The remainder of the paper is organized as follows. Section~\ref{sec:method}
introduces the sparse decomposition and the network architecture. The training
and sparsity-promoting procedures are then described, followed by supervised
and unsupervised benchmark problems in Section~\ref{sec:Benchmarks}. Finally, a discussion of the method is presented in Section~\ref{sec:Discussion}.

\section{Method}
\label{sec:method}
Many hyperelastic constitutive models have been proposed on the basis of physical arguments and experimental observations, see e.g., \cite{he_comparative_2022}. \rev{In this work we restrict ourselves to isotropic hyperelastic constitutive laws.} Representative examples include the \revv{Mooney-Rivlin}, polynomial, and Ciarlet--Geymonat models, whose strain-energy functions are summarized in \tabref{tab:model_comparison}. Although these models differ substantially in their functional forms and domains of applicability, developing and selecting an appropriate model traditionally requires  constitutive expertise. Moreover, a model that is effective for one material or loading regime may be unsuitable for another.

\begin{table}[htbp]
\centering
\caption{Some hyperelastic material models in literature.}
\label{tab:model_comparison}
\begin{tabular}{lc}
\toprule
Model & Strain energy function \\
\midrule
\rev{Mooney-Rivlin} & $\Psi = C_{10} (I_1 J^{-2/3} - 3) + C_{01} (I_2 J^{-4/3} - 3) + D (J-1)^2$ \\
Polynomial & $\Psi = \sum_{i, j = 0}^{n} C_{ij} (I_1- 3)^i (I_2-3)^j$  \\
Ciarlet-Geymonat & $\Psi = a (I_1-3) + b(I_2-3) + c(I_3^2-1) - d \ln(I_3)$ \\
\bottomrule
\end{tabular}
\end{table}

We propose an approach that learns a sparse decomposition of the strain energy from data with limited human intervention.
The method is based on the observation \rev{of the authors} that many hyperelastic constitutive models can be expressed as a sum of products of univariate functions of the invariants:
\begin{equation}
\Psi(\calI_\textrm{C}) = \sum_{m=1}^{N} a_{m}(I_1) b_{m}(I_2) c_{m}(J) \label{eq:SparseDecomposition}
\end{equation}
with the set of invariants $\calI_\textrm{C} = ( I_1, I_2, J )$ of the right Cauchy-Green tensor $\bfC = \bfF^T \cdot \bfF$ given as
\begin{equation}
  I_1 = \trC, \quad I_2 = \frac{1}{2} \left( \trC^2 - \tr[\bfC^2] \right), \quad J = \det(\bfF) = \sqrt{I_3}. 
\end{equation}
In general, many of the functions $a_{m}$, $b_{m}$, and $c_{m}$ are identically one, so only a small number of terms involve invariant couplings.
We use $J$ rather than the third invariant $I_3$ in the decomposition. The same decomposition strategy can also be formulated using $I_3$.

\rev{Based on the decomposition in Equation~\eqref{eq:SparseDecomposition}, we propose a suitable neural network architecture in Section~\ref{sec:NNArchitecture}. In this section, we also discuss how a zero-energy and stress-free reference configuration is enforced. Furthermore, we highlight how the proposed architecture guarantees convexity of the learned strain energy with respect to the invariants.}

\rev{The training procedure including a suitable loss function is described in Section~\ref{sec:Training}.
}

The decomposition in Equation~\eqref{eq:SparseDecomposition} most useful if only a small number of active functions, i.e., nonconstant functions, is identified. A conventional approach may represent the same energy accurately while distributing its response over many coupled subnetworks. Such a representation is generally non-unique and provides little information about which invariant dependencies are essential. Accuracy alone therefore does not guarantee an interpretable constitutive model. We address this issue by explicitly promoting sparsity, i.e., encouraging a small number of active functions. The sparsity-promoting techniques are discussed in Section~\ref{sec:Sparsity}.

\subsection{Neural network architecture}
\label{sec:NNArchitecture}
\rev{The decomposition of the strain energy in Equation~\eqref{eq:SparseDecomposition} is the starting point for the construction of the neural network architecture.}
Direct optimization of the functions $a_{m}$, $b_{m}$, and $c_{m}$ is challenging because they enter the energy multiplicatively. If one factor becomes small, the gradients of the remaining factors can vanish; if one factor becomes large, the resulting gradients can become unstable. In addition, the factorization has a scale indeterminacy: multiplying one factor by a constant and dividing another by the same constant leaves their product unchanged. These issues complicate both optimization and the interpretation of the learned factors.

We therefore use a log-domain parameterization. The subnetworks represent additive logarithmic factors, and the corresponding multiplicative contribution is recovered through an exponential. The strain energy density is then represented as
\begin{equation}
  \Psi(\calI_\textrm{C}) = \sum_{m=1}^{N}  \exp \left(\log(a_{m}(I_1)) + \log(b_{m}(I_2)) + \log(c_{m}(J)) \right).
\end{equation}
We further add a global scaling parameter $S > 1$ to the network 
\begin{equation}
  \Psi(\calI_\textrm{C}) = \sum_{m=1}^{N} S \exp \left(\frac{\Phi_m(\calI_\textrm{C})}{S} \right) 
\end{equation}
\revv{which scales the contribution of all modes simultaneously.}
The sum of all subnetworks of mode $m$ given as
\begin{equation}
  \Phi_m(\calI_\textrm{C}) = f_{m,1}(I_1) + f_{m,2}(I_2) + f_{m,3}(J) \label{eq:SubnetworkSum}
\end{equation}
where $f_{m,k}$ is the output of the subnetwork corresponding to the $k$-th input coordinate of mode $m$. \rev{For example, $f_{m, 2}(I_2)$ might learn an \revv{scaled} approximation of $\log(b_m(I_2))$.}
The scaling parameter is computed via
\begin{equation}
  S = \textrm{softplus}(s) + 1.0, \quad s \in \dsR
\end{equation}
where $s$ is an unconstrained parameter learned during training. 
The softplus function $\operatorname{softplus}(z) = \ln(1 + \exp(z))$ \rev{together with the factor $+1.0$} ensures that $S > 1$. 
\rev{The scaling parameter $S$ limits the magnitude of the exponential argument, thereby reducing the variation in the corresponding gradients. Without this scaling, the exponential derivative can vary by several orders of magnitude over the range of network inputs, which can destabilize training.}

\begin{figure}
  \def\svgwidth{18cm}
  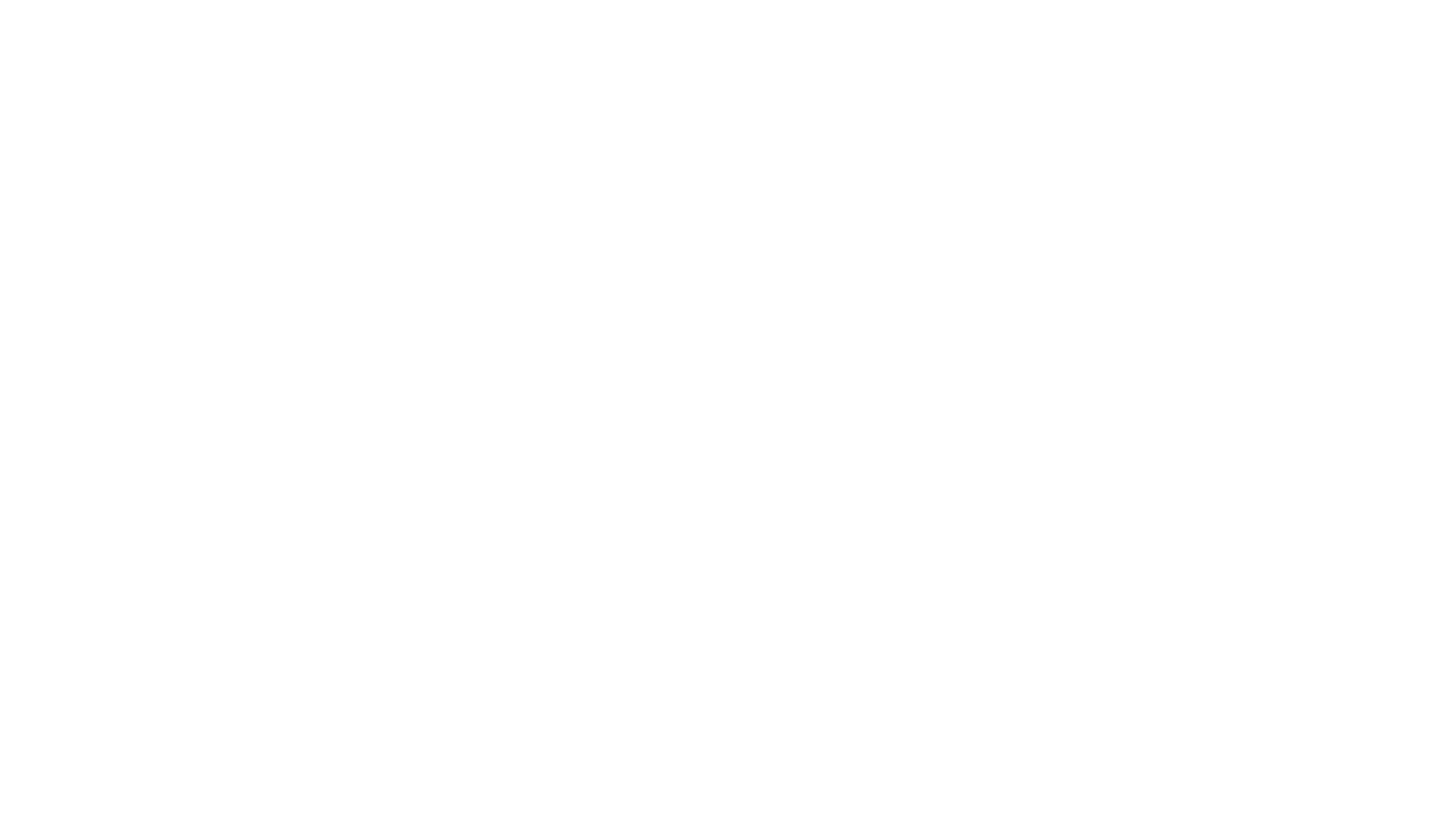
  \caption{Proposed network architecture. Each subnetwork $f_{m,k}$ is a one-dimensional input-convex neural network (ICNN) that takes a single invariant $I_k$ as input. The outputs of the subnetworks are summed to form $\Phi_m$, which is then exponentiated and scaled by $S$ to produce the contribution of mode $m$ to the strain energy. The total strain energy is obtained by summing over all modes. Sparsity is promoted through L1 regularization on the subnetwork outputs, intra-mode coupling penalties, and explicit mode pruning.}
  \label{fig:network_architecture}
\end{figure}

The total network architecture is illustrated in \figref{fig:network_architecture}. \rev{The network is made up of $m$ individual modes as described before. The last mode is dedicated exclusively to the volumetric part of the strain energy function. 
We use a trunk architecture, where the volumetric part is learned separately in the last mode. A similar split of the network has been used in \cite{holthusen_theory_2024}.}
The overall network architecture shares similarities with Tensor Neural Networks and Neural Additive Models \cite{wang_tensor_2024, agarwal_neural_2021}.

\subsubsection{Zero-energy and stress-free reference configuration}
We ensure zero energy at the reference state $\calI_0 := (I_1 = 3, I_2 = 3, J = 1)$ by subtracting the energy at the reference state from the energy at any other configuration. 
The energy at the reference state is given by
\begin{equation}
  \Psi_\mathrm{ref}(\calI_\textrm{0}) = \sum_{m=1}^{N} S \exp \left(\frac{\Phi_m(\calI_0)}{S} \right).
\end{equation}

The stress-free reference configuration is ensured by subtracting the hydrostatic part of the stress at the reference configuration from the stress at any other configuration.
The second Piola--Kirchhoff stress tensor can be expressed in terms of invariant derivatives as
\begin{equation}
\label{eq:stress_tensor_general}
\mathbf{S_\textrm{PK}} = 2 \frac{\partial \Psi}{\partial I_1} \mathbf{I} + 2 \frac{\partial \Psi}{\partial I_2} (I_1 \mathbf{I} - \mathbf{C}) + \frac{\partial \Psi}{\partial J} J \mathbf{C}^{-1}.
\end{equation}
Evaluating Equation~\eqref{eq:stress_tensor_general} at the reference configuration yields
\begin{equation}
\label{eq:S_ref_spherical}
\mathbf{S}_{\textrm{PK}, \mathrm{ref}} = \left( 2 \left.\frac{\partial \Psi}{\partial I_1}\right|_{\calI_0} + 4 \left.\frac{\partial \Psi}{\partial I_2}\right|_{\calI_0} + \left.\frac{\partial \Psi}{\partial J}\right|_{\calI_0} \right) \mathbf{I}.
\end{equation}
This reference stress is purely hydrostatic. The hydrostatic pressure $p_0$ at the reference configuration is defined as
\begin{equation}
\label{eq:p0_definition}
p_0 \equiv 2 \left.\frac{\partial \Psi}{\partial I_1}\right|_{\calI_0} + 4 \left.\frac{\partial \Psi}{\partial I_2}\right|_{\calI_0} + \left.\frac{\partial \Psi}{\partial J}\right|_{\calI_0}.
\end{equation}

Thus, the strain energy function is corrected by subtracting the hydrostatic pressure $p_0$ multiplied by the volumetric change $(J - 1)$, ensuring zero stress at the reference configuration.

The corrected energy predicted by the model is
\begin{equation}
\label{eq:psi_corrected}
\hat{\Psi}(\calI_\textrm{C}) = \Psi_{\mathrm{raw}}(\calI_\textrm{C}) - \Psi_{\mathrm{ref}} - p_0 (J - 1).
\end{equation}

\subsubsection{Architecture of the subnetworks}
Each one-dimensional subnetwork \rev{in Equation~\eqref{eq:SubnetworkSum}} is denoted by $f_{m,k}(x_k)$, where
$x_1=I_1$, $x_2=I_2$, and $x_3=J$. For $k \in \{1,2,3\}$, each subnetwork is
parameterized as a two-layer input-convex neural network followed by an
linear scaling neuron without bias.
The equation for a single subnetwork is given as
\begin{equation}
\label{eq:icnn_sub}
f_{m,k}(x_k) = W_{3,k}^{(m)} \cdot \left[ \bfW_{2,k}^{(m)} \cdot \operatorname{softplus}\left( \bfW_{1,k}^{(m)} x_k + \bfb_{1,k}^{(m)} \right) + b_{2,k}^{(m)} \right],
\end{equation}
where
\begin{itemize}
    \item $\bfW_{1,k}^{(m)} \in \dsR^{n_n \times 1}$, $\bfb_{1,k}^{(m)} \in \dsR^{10}$: unconstrained affine input layer;
    \item $\bfW_{2,k}^{(m)} \in \dsR^{1 \times n_n}_+$, $b_{2,k}^{(m)} \in \dsR$: non-negative hidden layer weights ($\bfW_{2,k}^{(m)} \ge 0$);
    \item $W_{3,k}^{(m)} \in \dsR_+$: non-negative mode scaling amplitude ($W_{3,k}^{(m)} \ge 0$).
\end{itemize}
Here, $n_n$ denotes the number of neurons in the hidden layer.
The softplus activation function $\operatorname{softplus}(z) = \ln(1 + e^z)$ is used to ensure convexity of the subnetwork. The non-negativity of the weights $\bfW_{2,k}^{(m)}$ and $W_{3,k}^{(m)}$ is enforced during training by a projection step after each gradient update. \revv{The linear scaling neuron $W_{3,k}^{(m)}$ and the global scaling parameter $S$ have very similar effects on the total response of the neural network. However, the global scaling parameter $S$ is shared across all modes, while the linear scaling neuron $W_{3,k}^{(m)}$ is specific to each subnetwork. Nevertheless, there are an infinite amount of combinations of $S$ and all $W_{3,k}^{(m)}$ of the modes that yield the same total response.}
The inclusion of the scaling factor $W_{3,k}^{(m)}$ allows for simple control of the sparsity. To avoid scale indeterminacy, we enforce a unit Euclidean norm constraint on the row vector \rev{$\bfW_{2,k}^{(m)}$, i.e.,
\begin{equation}
\norm{\bfW_{2,k}^{(m)}}_2
= \left(\sum_{j=1}^{n_n} \left(W_{2,k}^{(m)}\right)_j^2\right)^{1/2}
= 1.
\end{equation}}
This is done by rescaling $\bfW_{2,k}^{(m)}$ after each gradient update.

\subsubsection{Convexity}
Using input-convex subnetworks with log-domain training ensures that the learned strain energy function $\Psi(I_1, I_2, J)$ is convex in the set of invariants $\calI_\textrm{C} \in \dsR^3_+$.

\revv{The sum of the individual subnetworks $\Phi_m$ of mode $m$ is a sum of convex functions in the disjoint coordinates $I_1, I_2, J$, hence jointly convex in the set of invariants} $\calI_\textrm{C}$.
The total contribution of each mode $m$, $g(u) = S(\exp({\Phi_m/S}) - exp(\Phi_m(\calI_0)/S))$ has first and second derivatives:
\begin{equation}
g'(\Phi_m) = \exp({\Phi_m/S}) > 0, \quad g''(\Phi_m) = \frac{1}{S} \exp({\Phi_m/S}) > 0 \quad (\text{since } S > 1).
\end{equation}
Thus, $g$ is convex and non-decreasing. The composition of a convex non-decreasing function with a convex function is convex, thus $\Psi_m(\calI_\textrm{C}) = g(\Phi_m(\calI_\textrm{C}))$ is convex in $\calI_\textrm{C}$.
Finally, $\Psi(\calI_\textrm{C}) = \sum_{m=1}^M \Psi_m(\calI_\textrm{C})$ is a sum of convex functions, ensuring global convexity of $\Psi(\calI_\textrm{C})$.

\subsubsection{Initialization}
The network is initialized using Glorot uniform initialization for the weights $\bfW_{1,k}^{(m)}$, $\bfW_{2,k}^{(m)}$ and $\bfW_{3,k}^{(m)}$, and zero initialization for the biases $\bfb_{1,k}^{(m)}$ and $b_{2,k}^{(m)}$. To ensure positive weights, the absolute value of all weights is used. To break symmetry, some of the weights $W_{3,k}^{(m)}$ are scaled by a factor of $0.05$. \revv{The scaled weights are selected deliberately: two weights are scaled in the first modes, whereas only one is scaled in the subsequent modes, ensuring that every combination of scaled parameters occurs at least once.}
The last mode is dedicated exclusively to the volumetric part of the strain energy function, i.e., $W_{3, 3}^{(\textrm{last})} = W_{3, 2}^{(\textrm{last})} = 0$. It is initialized to produce a symmetric U-shaped function. \rev{In particular, the weights $\left(W_{1, 3}^{(\textrm{last})}\right)_j$ are initialized with pair-wise opposite sign slopes. The biases are chosen as $\bfb_{1, 3}^{(\textrm{last})} = \bfW_{1, 3}^{(\textrm{last})}$.}

\subsection{Training}
\label{sec:Training}
\revv{In this contribution, we investigate both supervised and unsupervised training. In the supervised setting, the network is trained directly on labeled data consisting of deformation invariants and the corresponding strain energy values. The loss therefore measures the pointwise discrepancy between the predicted and reference energies. In the unsupervised setting, no local strain energy or stress labels are provided. Instead, the network is trained from measured deformation fields and global reaction forces by enforcing equilibrium and agreement with the measured forces. 
The details are explained in the next two subsections.}

\subsubsection{Supervised training}
\revv{For supervised training in the log-domain, we use the Poisson negative log-likelihood as loss} defined as
\begin{equation}
\label{eq:poisson_loss}
\calL_{\text{Poisson}}(\bfy, \hat{\bfy}) = \frac{1}{N} \sum_{i=1}^N \left[ \hat{y}_i - y_i \ln(\hat{y}_i) \right], 
\end{equation}
\rev{where $y_i$ are the true values for the strain energy function and $\hat{y}_i$ are the predicted values for the strain energy function as} 
\begin{equation}
  \hat{y}_i = \max(\hat{\Psi}(\calI_\textrm{C}), \, \epsilon_{\text{safe}}),
\end{equation}
with the numerical parameter $\epsilon_{\text{safe}} = 10^{-4}$ ensuring strict positivity of the predicted values. 
\revv{It should be remarked, that the Poisson negative log-likelihood is not zero for perfect predictions, but it is minimized when the predicted values match the true values.}
The Poisson negative log-likelihood has several advantages.
The gradient of $\calL_{\text{Poisson}}$ with respect to the predicted values is
\begin{equation}
\frac{\partial \calL_{\text{Poisson}}}{\partial \hat{y}_i} = 1 - \frac{y_i}{\hat{y}_i} = \frac{\hat{y}_i - y_i}{\hat{y}_i}. \label{eq:PoissonLoss}
\end{equation}
This gradient therefore scales with the relative error.
Therefore, errors in large and small energy regions are treated equally, which is important for the training of hyperelastic material models, where the strain energy can vary over several orders of magnitude.

The gradient with respect to the internal weights for a single active mode is given as
\begin{equation}
\frac{\partial \calL}{\partial \Phi_m} = \frac{\partial \calL}{\partial \hat{y}} \cdot \frac{\partial \hat{y}}{\partial \Phi_m} = \frac{\partial \calL}{\partial \hat{y}} \cdot \exp\left(\frac{\Phi_m}{S}\right).
\end{equation}
\rev{In this expression we neglected the contributions from the reference-state energy and stress corrections to present the general behavior of the gradient more concisely \revv{but it is taken into account in practice}. The network is trained on the corrected energy $\hat{\Psi}$, which includes these corrections.}
By using the Poisson loss, the gradient with respect to the internal weights becomes
\begin{equation}
\frac{\partial \calL_{\text{Poisson}}}{\partial \Phi_m} = \left(\frac{\hat{y} - y}{\hat{y}}\right) \cdot \exp\left(\frac{\Phi_m}{S}\right) \approx \left(\frac{\hat{y} - y}{\hat{y}}\right) \cdot  \frac{\hat{y}}{S} = \frac{1}{S} \left( \hat{y} - y \right)
\end{equation}
For a single active mode, the gradient with respect to the mode output is therefore linear in the prediction error.

By comparison, using a standard mean squared error (MSE) loss gives
\begin{equation}
\frac{\partial \calL_{\text{MSE}}}{\partial \Phi_m} = 2(\hat{y} - y) \cdot \exp\left(\frac{\Phi_m}{S}\right) \propto 2(\hat{y} - y) \cdot \hat{y} \sim \calO(\hat{y}^2),
\end{equation}
which scales quadratically in the predicted values $\hat{y}$.
This can lead to catastrophic gradient explosion and numerical overflow during backpropagation.

During training, all modes are trained simultaneously using the Adam optimizer. The positivity of the weights $\bfW_{2,k}^{(m)}$ and $W_{3,k}^{(m)}$ is enforced by a projection step after each gradient update as 
\begin{align}
\bfW_{2,k}^{(m)} &\leftarrow \max\left(0, \, \bfW_{2,k}^{(m)}\right), \label{eq:proj_w2} \\
W_{3,k}^{(m)} &\leftarrow \max\left(0, \, W_{3,k}^{(m)} \right). \label{eq:proj_w3}
\end{align}

In the last training stage, we switch to the L-BFGS-B optimizer for fine-tuning of the network parameters. The L-BFGS-B optimizer allows the incorporation of the non-negativity constraints of the weights $\bfW_{2,k}^{(m)}$ and $W_{3,k}^{(m)}$ directly into the optimization problem. The L-BFGS-B optimizer is a quasi-Newton method that approximates the inverse Hessian matrix using a limited amount of memory.

\subsubsection{Unsupervised training}
\revv{
For the unsupervised training, we assume that the deformation fields and global reaction forces are measured experimentally. The network is trained to minimize the discrepancy between the predicted and measured reaction forces while enforcing equilibrium at unconstrained degrees of freedom. The identification follows the EUCLID framework \cite{thakolkaran_nn-euclid_2022}. }

\revv{
The training objective combines an equilibrium residual with a reaction-force
residual. Let $\mathbf{r}_{\mathrm{free}}$ denote the internal force vector
restricted to unconstrained degrees of freedom, and let $F_{\mathrm{pred}}$
and $F_{\mathrm{meas}}$ denote the predicted and measured reaction forces,
respectively. For each selected load step, the loss term is defined as
\begin{equation}
\mathcal{L}_{\mathrm{EUCLID}} =
\frac{\lVert\mathbf{r}_{\mathrm{free}}\rVert_2^2}
  {F_{\mathrm{meas}}^2}
+ \lambda_R \left(\frac{F_{\mathrm{pred}}-F_{\mathrm{meas}}}
  {F_{\mathrm{meas}}}\right)^2, \label{eq:EUCLIDLoss}
\end{equation}
where $\lambda_R=50$ weights the global reaction-force constraint. The first
term enforces weak equilibrium in the measured deformation state, whereas the
second term ensures agreement with the experimentally available resultant
force. The objective is averaged over the selected load steps and augmented
with the $L_1$ regularization and the intra-mode coupling penalty described
in Section~\ref{sec:Sparsity}.
}

\revv{The same optimizer setup using Adam and L-BFGS-B as for the supervised training are used.}

\subsection{Sparsity}
\label{sec:Sparsity}
We promote sparsity by three mechanisms: L1 regularization, intra-mode coupling penalties, and explicit mode pruning. 

Spurious modes with small amplitudes can hinder interpretability. Therefore, we introduce an $L_1$ penalty on the amplitudes $W_{3,k}^{(m)}$ to encourage exact zeros.
The $L_1$ regularization term is defined as
\begin{equation}
\calR_{\text{L1}} = \sum_{m=1}^{M} \sum_{k=1}^{3} \left\lvert W_{3,k}^{(m)} \right\rvert.
\end{equation}
This term is weighted by a factor of $\lambda_{\mathrm{L1}}$ and added to the loss. \rev{The $L_1$ penalty also drives amplitudes toward zero, suppressing subnetworks with small contributions to the strain energy. To ensure that sufficiently small amplitudes are set exactly to zero, we use a proximal update after each gradient step in the form}
\begin{equation}
W_{3,k}^{(m)} \leftarrow \max\left(0, W_{3,k}^{(m)} - \lambda_{\mathrm{L1}}\eta\right),
\end{equation}
where $\eta$ is the step size \rev{of the optimizer}. 
If $W_{3,k}^{(m)}=0$, then $f_{m,k}(I_k)=0$ and the corresponding factor in the exponential representation is one. Each inactive subnetwork contributes no deformation-dependent energy and cannot generate a spurious mode.

Furthermore, we penalize modes that simultaneously depend on all three invariants. Such three-way couplings are often difficult to interpret, whereas univariate and bivariate contributions can be related more directly to established constitutive terms. We therefore introduce the intra-mode coupling regularizer
\begin{equation}
\calR_{\text{3 cross}} = \sum_{m=1}^{M} W_{3,1}^{(m)} W_{3,2}^{(m)} W_{3,3}^{(m)}.
\end{equation}
Because $W_{3,k}^{(m)} \geq 0$, the contribution of a mode to this penalty vanishes if and only if at least one of its three subnetworks is inactive. The regularizer therefore permits arbitrary univariate modes, such as modes depending only on $I_1$, $I_2$, or $J$, as well as bivariate couplings, while discouraging three-way couplings of $I_1$, $I_2$, and $J$. For models with two invariants, i.e., incompressible materials, the analogous product of the two amplitudes defines $\calR_{\text{2 cross}}$. The corresponding terms enter the objective with factors $\lambda_{\text{2 cross}}$ and $\lambda_{\text{3 cross}}$, respectively. 

Finally, redundant modes are removed during training. \revv{Two modes are considered redundant if they depend on the same set of invariants and both make non-constant contributions. In this case, we remove the mode whose contribution to the total strain energy, measured over the training domain, is smaller. Although pruning temporarily increases the loss, the remaining mode can subsequently adapt to compensate for the removed contribution. If the modes depend on two invariants, pruning one of them restricts the model to a single rank-one approximation of the corresponding two-dimensional contribution and may therefore discard relevant information. Nevertheless, we allow this pruning because such rank-one representations are commonly used in hand-enginneered constitutive models.}

As an alternative, we also investigated a regularization that directly penalizes similarity between the learned one-dimensional functions of different modes via the cosine similarity. This approach was not successful: the modes compete to reduce the similarity penalty, which can lead to unstable redistribution of contributions and spurious terms. 

\revv{In total, the training loss in supervised as well as unsupervised training are augmented by the sparsity-promoting terms given as
\begin{equation}
  \calL_{\textrm{Sparsity}} = \lambda_{\textrm{L1}} \calR_{\textrm{L1}} + \lambda_{\textrm{2 cross}} \calR_{\textrm{2 cross}} + \lambda_{\textrm{3 cross}} \calR_{\textrm{3 cross}}
\end{equation}
}

\section{Benchmarks}
\label{sec:Benchmarks}
We evaluate the method in two supervised benchmarks and one unsupervised benchmark. In the supervised learning tasks \revv{in Sections~\ref{sec:PolyMat} and \ref{sec:MooneyRivlin}}, we generate a dataset of the strain energy for a given deformation from a known material model and train the network to learn the underlying strain energy function. In the unsupervised learning task \revv{in Section~\ref{sec:Unsupervised}}, we train the network on synthetic experimental data without any prior knowledge of the underlying material model. \rev{In each benchmark, we use a different material law to identify in order to showcase the applicability of the method to a variety of constitutive laws.}

\subsection{Polynomial material model}
\label{sec:PolyMat}
We consider an isotropic and incompressible material with the strain energy function
\begin{equation}
\Psi = C_{20} (I_1 - 3)^2 + C_{01} (I_2 - 3) 
\end{equation}
with the material parameters $C_{20} = \SI{0.5}{MPa}$ and $C_{01} = \SI{0.2}{MPa}$. \rev{This model is a special case of the general polynomial material model presented in Table \ref{tab:model_comparison}.}
We generate data along three standard loading paths: uniaxial tension, equi-biaxial tension, and pure shear. For an incompressible isotropic material, the uniaxial and equi-biaxial paths form the lower and upper boundaries of the admissible region in the $(I_1,I_2)$ plane, respectively, while pure shear provides an additional path through its interior.

The stretch ratio $\lambda = L/L_0$ measures the current length $L$ relative to the reference length $L_0$. Thus, $\lambda=1$ denotes the undeformed configuration and $\lambda>1$ denotes tensile extension. We use 50 data points for each loading path, resulting in 150 training points in total. The corresponding stretch and invariant ranges are listed in \tabref{tab:benchmark_protocols}.

\begin{table}[htbp]
\centering
\caption{Benchmark loading for the identification of the polynomial material model.}
\label{tab:benchmark_protocols}
\begin{tabular}{l|lll}
\toprule
Loading mode & Stretch range & $I_1$ range & $I_2$ range \\
\midrule
Uniaxial & [0.6, 3.0] & [3.0, 9.7] & [3.0, 6.1] \\
Pure shear & [0.6, 2.5] & [3.0, 7.4] & [3.0, 7.4] \\
Equi-biaxial & [0.6, 2.0] & [3.0, 8.1] & [3.0, 16.5] \\
\bottomrule
\end{tabular}
\end{table}

The training details are summarized in \tabref{tab:TrainingDetails}. 
The network is initialized with 5 modes, each consisting of 2 subnetworks for the invariants $I_1$ and $I_2$. As the material is incompressible, the subnetwork for the invariant $J$ is not used.
Each subnetwork consists of 20 hidden neurons, resulting in a total of 640 trainable parameters. 

\begin{table}[htbp]
\centering
\caption{Training details for the identification of the polynomial material model.}
\label{tab:TrainingDetails}
\begin{tabular}{ll}
\toprule
Learning rate & $1 \times 10^{-3}$ \\
Optimizer & Adam until 25000 epochs \\
& L-BFGS for the last 5000 epochs \\
Total number of epochs & 30000 \\
Regularization & $\lambda_\textrm{L1} = \lambda_\textrm{2 cross}= 1 \times 10^{-3}$ \\
& \rev{linear} ramp up until 5000 epochs \\
Mode pruning & After 10000 epochs \\
\bottomrule
\end{tabular}
\end{table}

\begin{figure}
\centering
\includegraphics[width=0.5\textwidth]{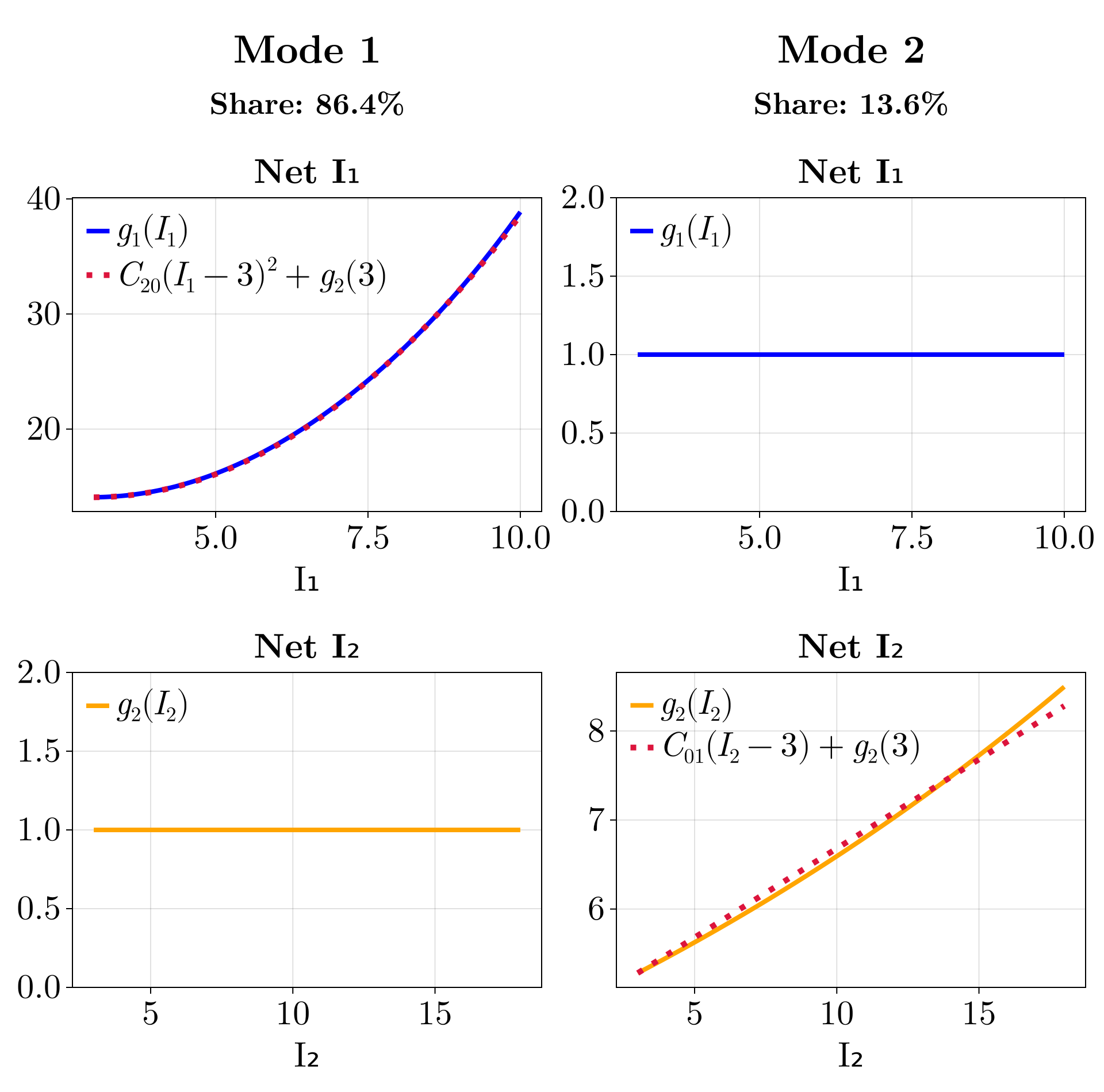}
\caption{Identified material model for the polynomial material model.}
\label{D1_Architecture}
\end{figure}

The final network after training is presented in Figure~\ref{D1_Architecture}. Columns represent modes, and rows represent subnetworks for $I_1$ and $I_2$. The reference-state energy and stress corrections are omitted from the figure. \rev{For each mode, the relative contribution to the total strain energy is indicated.}
The network contains two active modes: one depends only on $I_1$, and the other only on $I_2$. The corresponding analytical functions are shown as dotted red lines.
The learned model recovers the sparse structure of the polynomial model and accurately approximates its strain energy. Training takes a few minutes on a standard desktop computer.

Despite the ill-posed nature of sparse constitutive identification, the network recovers the number of active modes and the functional form of the reference energy in this benchmark.

\begin{figure}
\centering
\includegraphics[width=0.8\textwidth]{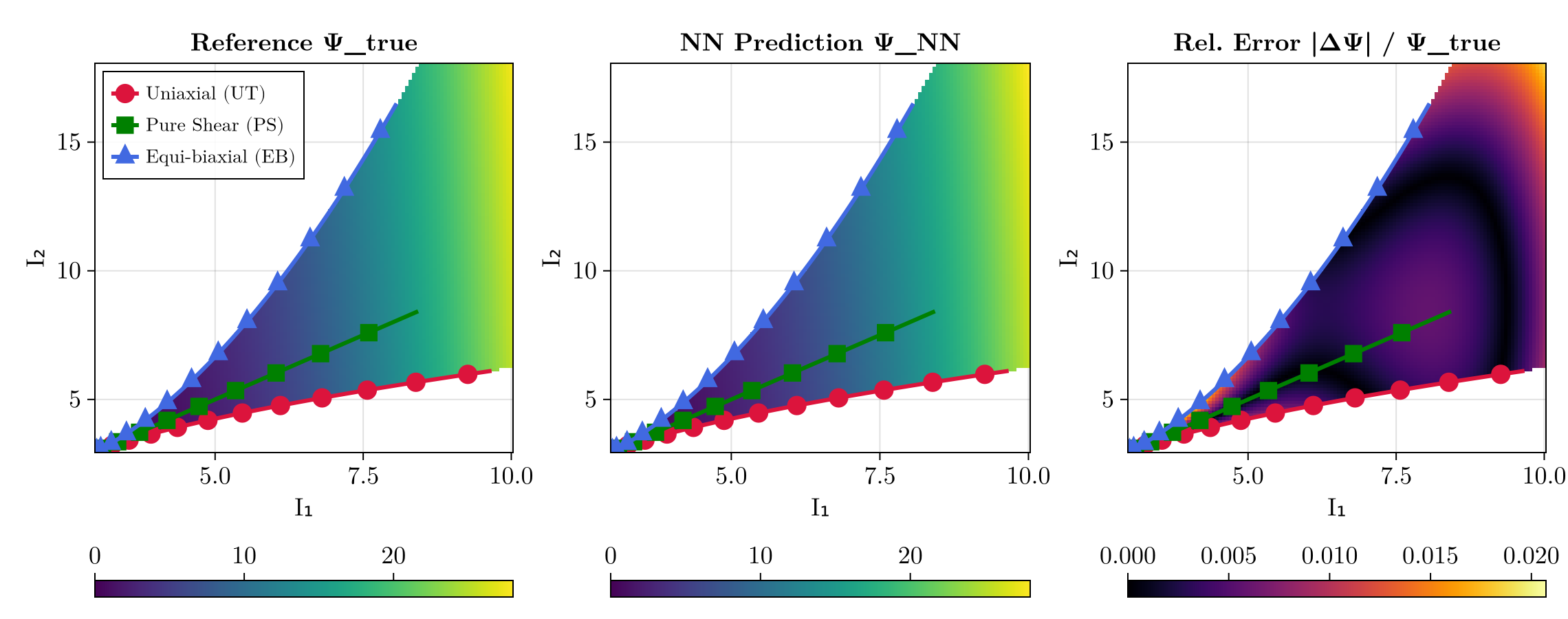}
\caption{Comparison of the learned strain energy function with the true strain energy function for the polynomial material model. Loading modes for the training data are shown as colored lines.}
\label{D1_Energy}
\end{figure}

The generalization ability of the network is demonstrated in Figure~\ref{D1_Energy}, where the learned strain energy function is visualized in the invariant plane. The learned model accurately captures the energy landscape of the polynomial model, even in regions of the invariant space that were not included in the training data. The largest error occurs on the right side of the plot with a high value of $I_1$ larger than any training data point. This highlights the ability of the network to interpolate between training data and generalize to unseen deformation states.

\begin{figure}
\centering
\includegraphics[width=0.5\textwidth]{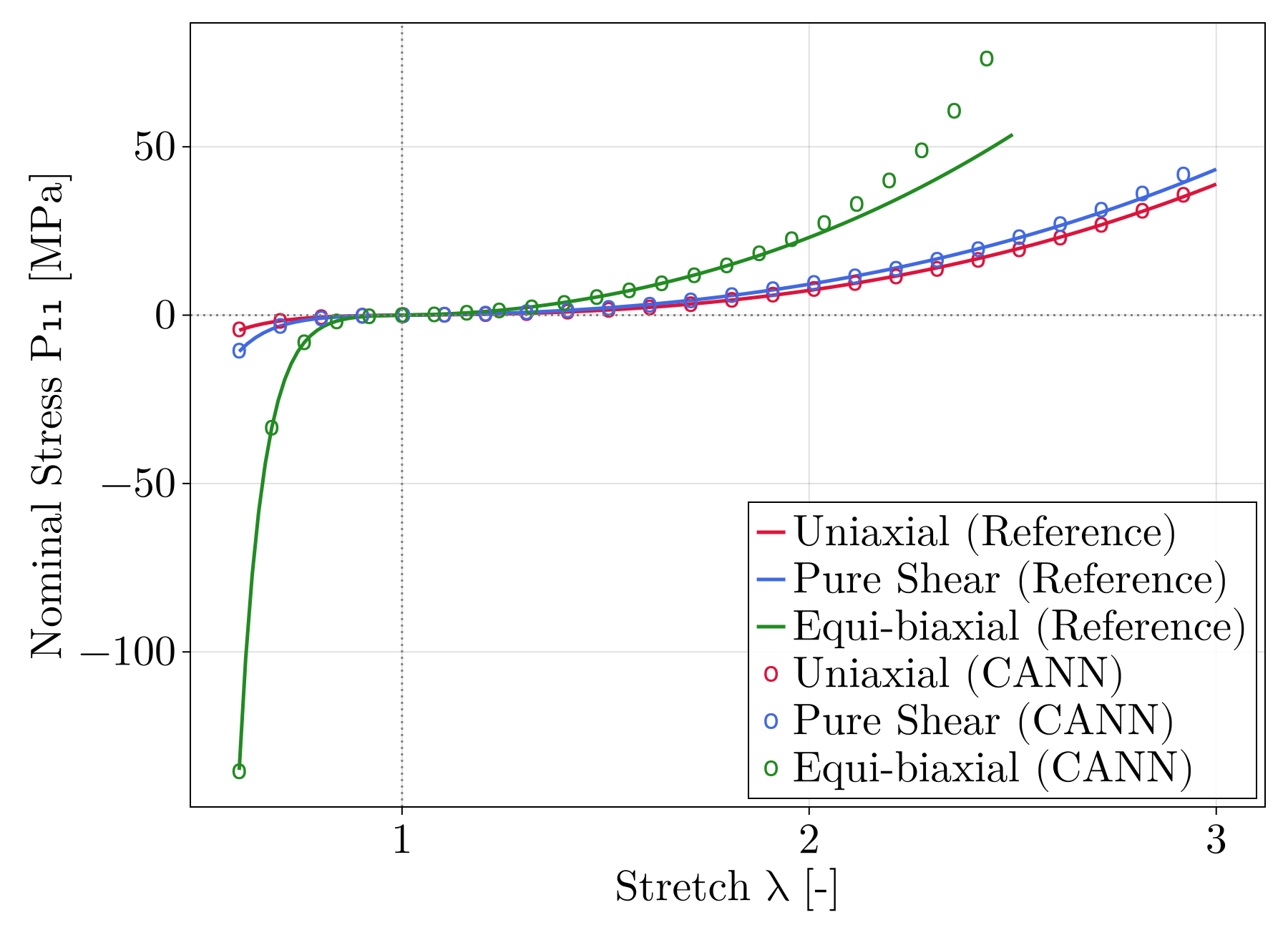}
\caption{Loading modes for the polynomial material model.}
\label{D1_Loading}
\end{figure}

In Figure~\ref{D1_Loading}, the learned model is evaluated under the three loading modes used for training. The predicted stress-strain curves are compared to the ground truth data, showing close agreement between the learned model and the true material response for the uniaxial and pure shear loading conditions. For the equi-biaxial loading, the learned model overestimates the stress at high stretch ratios, which is consistent with the error observed in the energy landscape. This region corresponds to high values of $I_1$ and $I_2$.

\begin{figure}
\centering
\includegraphics[width=0.5\textwidth]{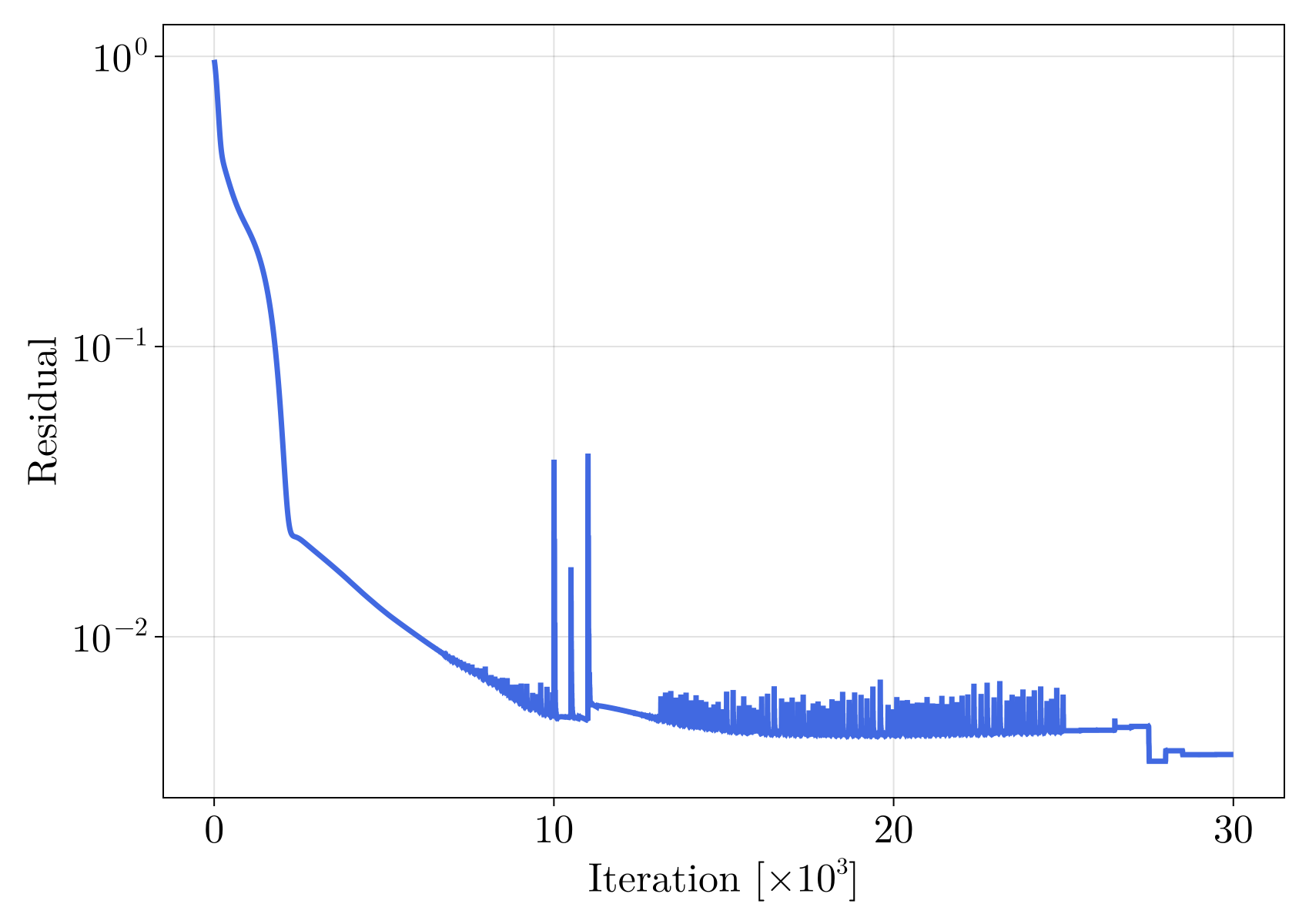}
\caption{Convergence history during training of the neural network on the polynomial material model.}
\label{D1_Convergence}
\end{figure}

Finally, in Figure~\ref{D1_Convergence}, the convergence history of the training process is shown. \rev{The residual is defined as the relative $L_2$ error norm between the networks prediction and the ground-truth reference strain energy.}
The \rev{residual} decreases rapidly during the initial stages of training, followed by slower convergence as the network approaches the optimal solution. Around 10000 epochs, the mode pruning is activated, leading to a spike in the training loss. Nevertheless, the network is able to recover soon after and reduce the training loss even further.
The final loss value of $0.4\%$ indicates that the learned model achieves a high level of accuracy in approximating the underlying strain energy function.

The effect of the sparse regularization is important for the identification of the underlying material model. Without regularization, the network may learn a different material model with more active modes, which is less interpretable and harder to analyze. The effect of the regularization is discussed in Section~\ref{sec:SparseReg}.
We also investigated the effect of the network size and dataset size. For a sufficiently large network with more than 10 hidden neurons per subnetwork, the network is able to robustly identify the underlying material model for different random initializations. A similar effect is observed for the dataset size. For a sufficiently large dataset with more than approximately 30 training points per loading path, the network is able to robustly identify the underlying material model for different random initializations. However, for a lower number of training points, the network may learn a different material model. This robustness and the alternatively learned material model are discussed in Section~\ref{sec:Noise}.

\subsubsection{Influence of sparsity regularization}
\label{sec:SparseReg}
We investigate the influence of the sparsity regularization on the identification of the underlying material model.
The same training procedure and hyperparameters as in Section~\ref{sec:PolyMat} are used, but the sparsity regularization is varied. We showcase the identified material model for three different regularization weights $\lambda_\textrm{L1} = \lambda_\textrm{2 cross} \in \{0, 1 \times 10^{-3}, 1 \times 10^{-1}\}$ and present the mean error on the training and an additional test dataset across three different runs. Here, by error, the root mean square error is meant. The test dataset consists of 1000 points in the admissible part of the invariant plane, with $I_1 \in [3.0, 10.0]$ and $I_2 \in [3.0, 16.0]$. In this study, we vary the regularization weight $\lambda_\textrm{L1}$ and $\lambda_\textrm{2 cross}$ simultaneously, i.e., $\lambda_\textrm{L1} = \lambda_\textrm{2 cross}$.
Mode pruning is disabled in this investigation. Only active modes containing at least one nonconstant function are shown.

\begin{figure}
\centering
\includegraphics[width=0.8\textwidth]{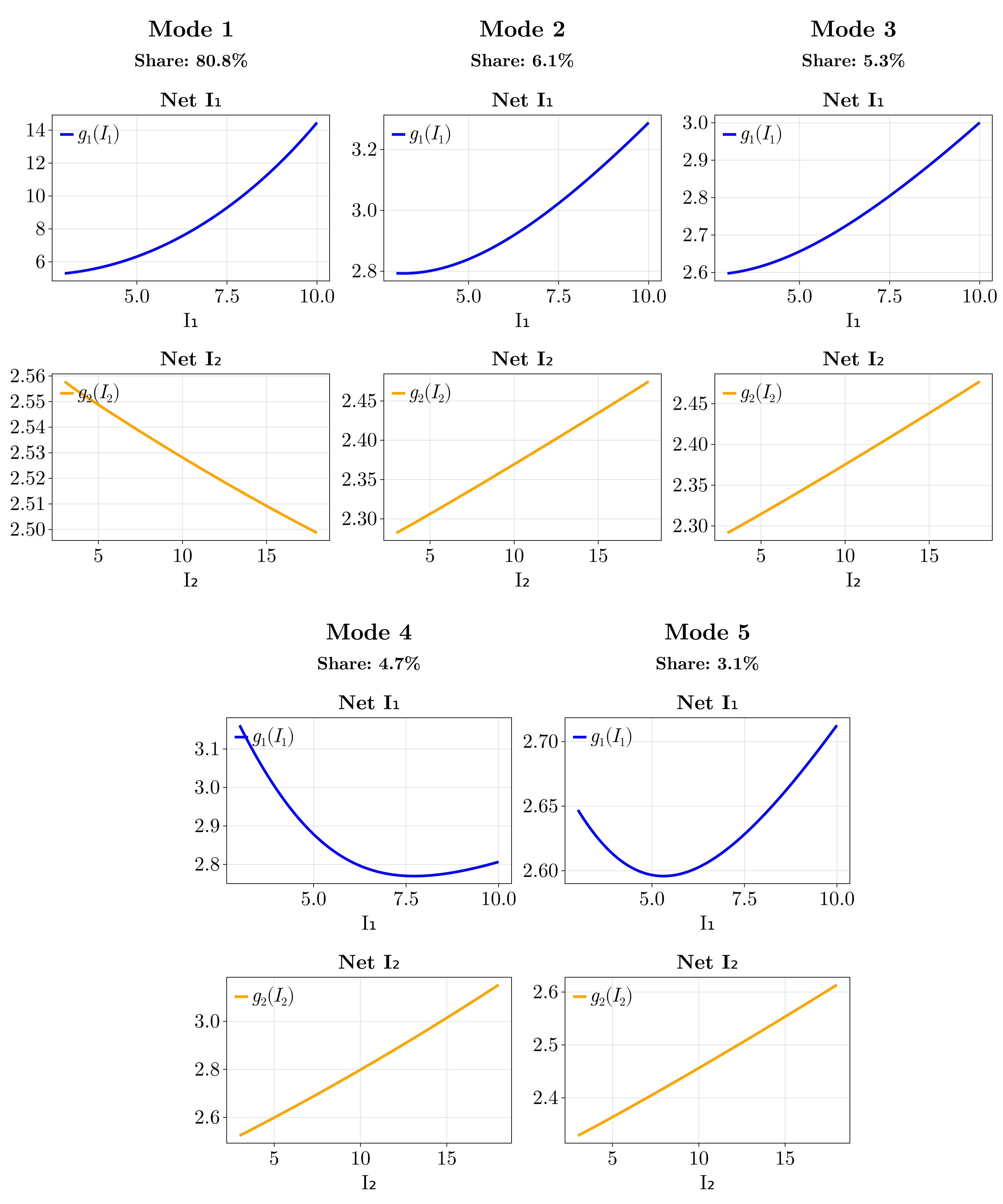}
\caption{Unregularized identification: Identified material model for the polynomial material model with $\lambda_\textrm{L1} = \lambda_\textrm{2 cross} = 0$.}
\label{fig:D1_Regularization_0}
\end{figure}

First, we consider training without regularization, i.e., $\lambda_\textrm{L1} = \lambda_\textrm{2 cross} = 0$, is investigated. The identified material model is shown in Figure~\ref{fig:D1_Regularization_0}. The network identifies a material model with 5 coupled and active modes. \rev{The modes are ordered by their contribution to the total strain energy.} The learned network has a very low error on training and test data of 0.13\% and 0.26\%, respectively. However, the learned material model is complicated to interpret as all individual modes are coupled in $I_1$ and $I_2$. For multiple networks, the contribution is nearly constant with only limited change across the input.

\begin{figure}
\centering
\includegraphics[width=0.5\textwidth]{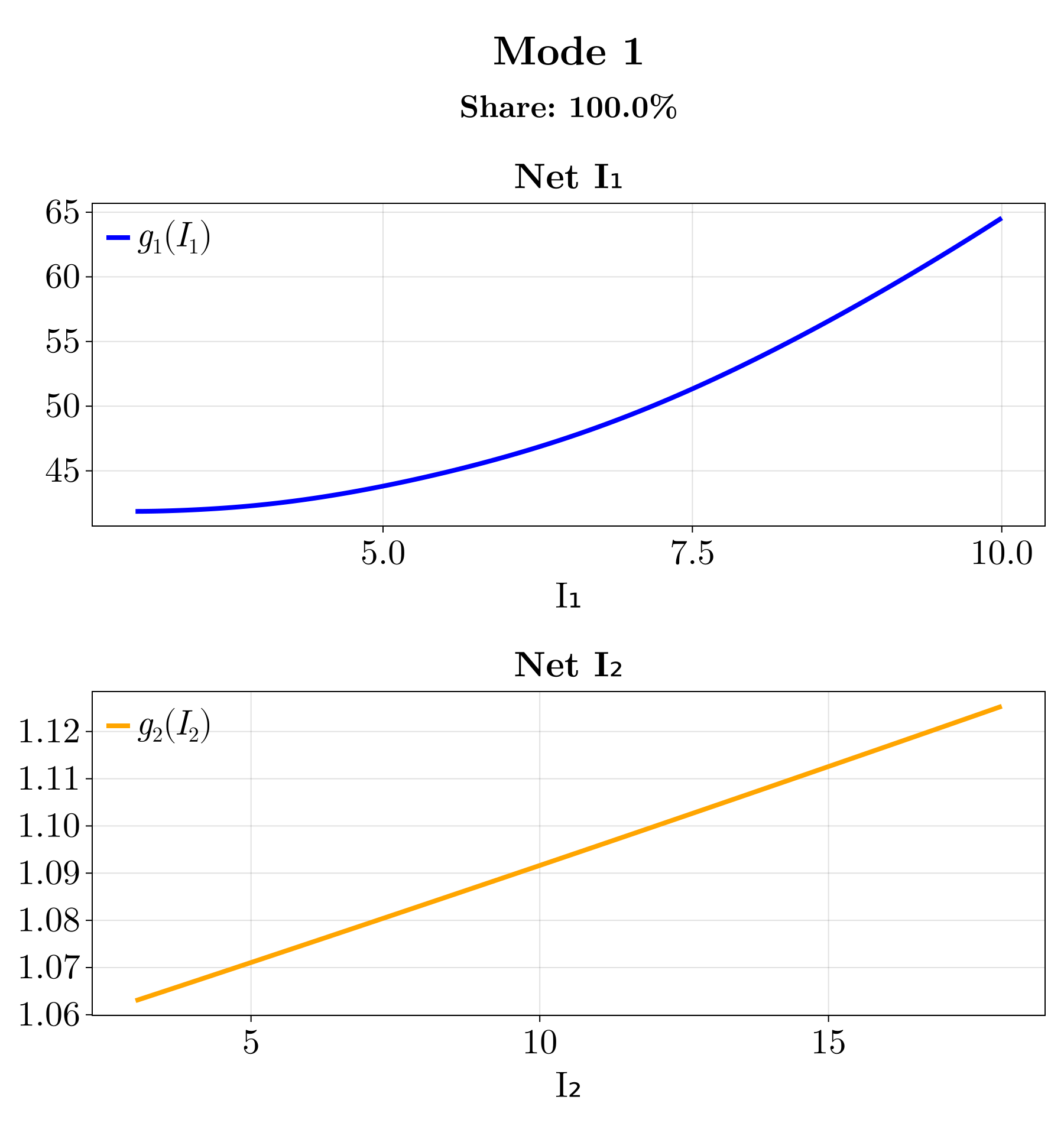}
\caption{Strong regularization: Identified material model for the polynomial material model with $\lambda_\textrm{L1} = \lambda_\textrm{2 cross} = 1 \times 10^{-1}$.}
\label{fig:D1_Regularization_1e-3}
\end{figure}

Second, a high regularization weight of $\lambda_\textrm{L1} = \lambda_\textrm{2 cross} = 1 \times 10^{-1}$ is used. The identified material model is shown in Figure~\ref{fig:D1_Regularization_1e-3}. The network identifies a material model with 1 active mode. Strong regularization leads to a sparser material model with fewer active modes, which is more interpretable and easier to analyze. However, the training and test errors increase to 0.54\% and 1.4\%, respectively.

\begin{figure}
\centering
\includegraphics[width=0.8\textwidth]{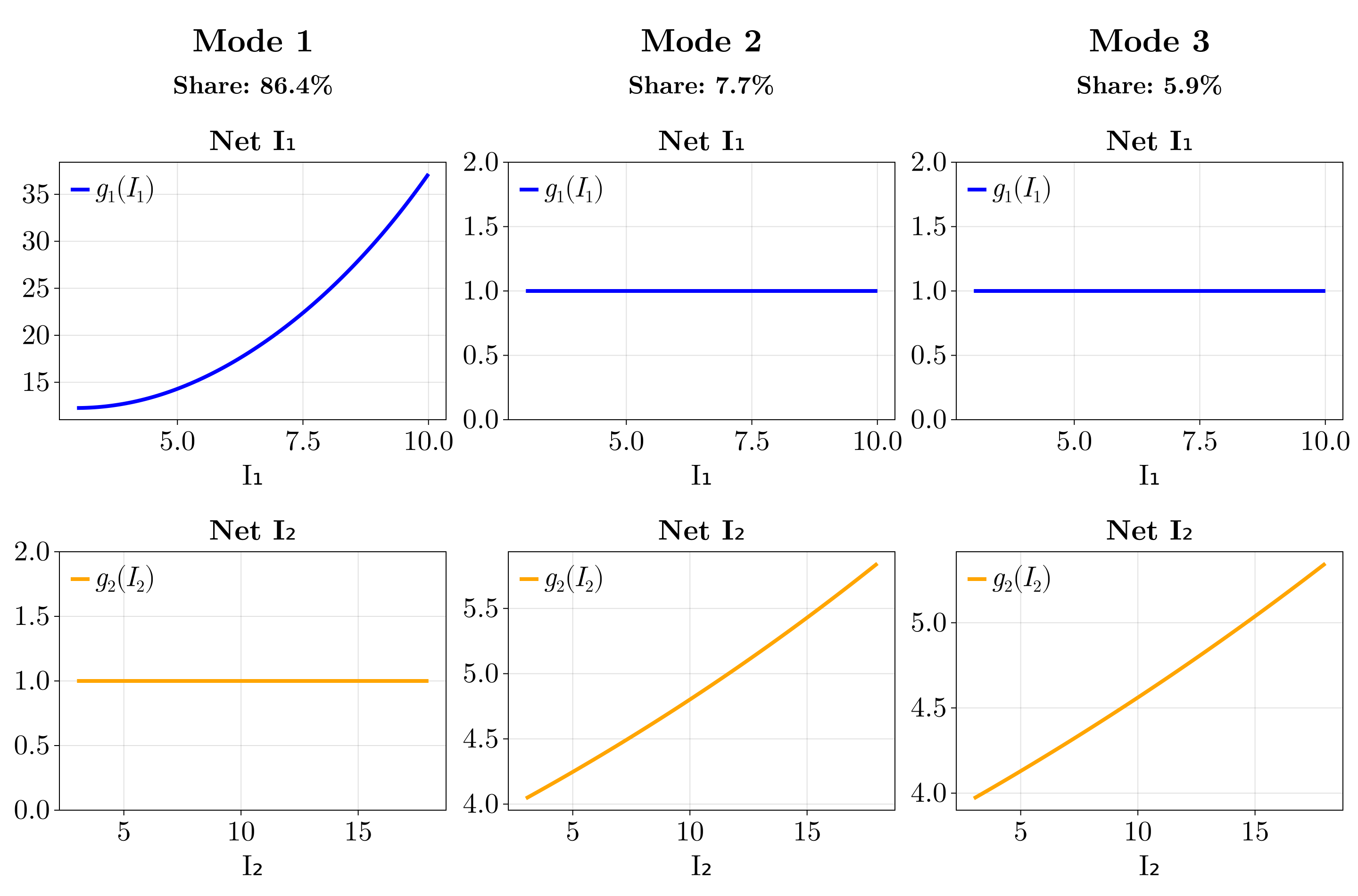}
\caption{Medium regularization: Identified material model for the polynomial material model with $\lambda_\textrm{L1} = \lambda_\textrm{2 cross} = 1 \times 10^{-3}$ without mode pruning.}
\label{fig:D1_Regularization_1e-1}
\end{figure}

Finally, a regularization weight of $\lambda_\textrm{L1} = \lambda_\textrm{2 cross} = 1 \times 10^{-3}$ is used as in the original study. Without mode pruning, the network retains three active modes: one depends only on $I_1$, and two depend only on $I_2$. The identified material model is shown in Figure~\ref{fig:D1_Regularization_1e-1}. The error on the training and test data is 0.44\% and 0.63\%, respectively, providing a compromise between accuracy and interpretability.

The results show that the sparsity regularization has a significant influence on the identification of the underlying material model. A larger regularization weight leads to a sparser material model with fewer active modes, which \rev{can be easier to analyze}. However, an excessively large regularization weight may lead to an oversimplified material model that does not capture the underlying material behavior accurately. Therefore, a careful selection of the regularization weight is crucial for the identification of a meaningful and interpretable material model. Classical model selection techniques may be used to select the optimal regularization weight, e.g., as done in \cite{urreaquintero_automated_2026}. However, \revv{in this paper}, a small regularization weight \rev{in the order $\calO(1\times10^{-3})$} is often sufficient to achieve a good balance between accuracy and interpretability.

\subsubsection{Noise}
\label{sec:Noise}
Next, we investigate the effect of noise in the training data on the identification procedure.
We add Gaussian noise with a standard deviation given as a percentage of the maximum value of the strain energy function. \rev{The data is clipped to positive values.} The same training procedure and hyperparameters as in Section~\ref{sec:PolyMat} are used. For each noise level, we train five networks with the same architecture and different random initializations to assess the robustness of the identification procedure.
For evaluation, a test set with 1000 points is generated in the admissible part of the invariant plane, with $I_1 \in [3.0, 10.0]$ and $I_2 \in [3.0, 16.0]$. The test set is used to evaluate the performance of the learned model on unseen data and to assess the generalization ability of the network. 

\begin{figure}
\centering
\includegraphics[width=0.5\textwidth]{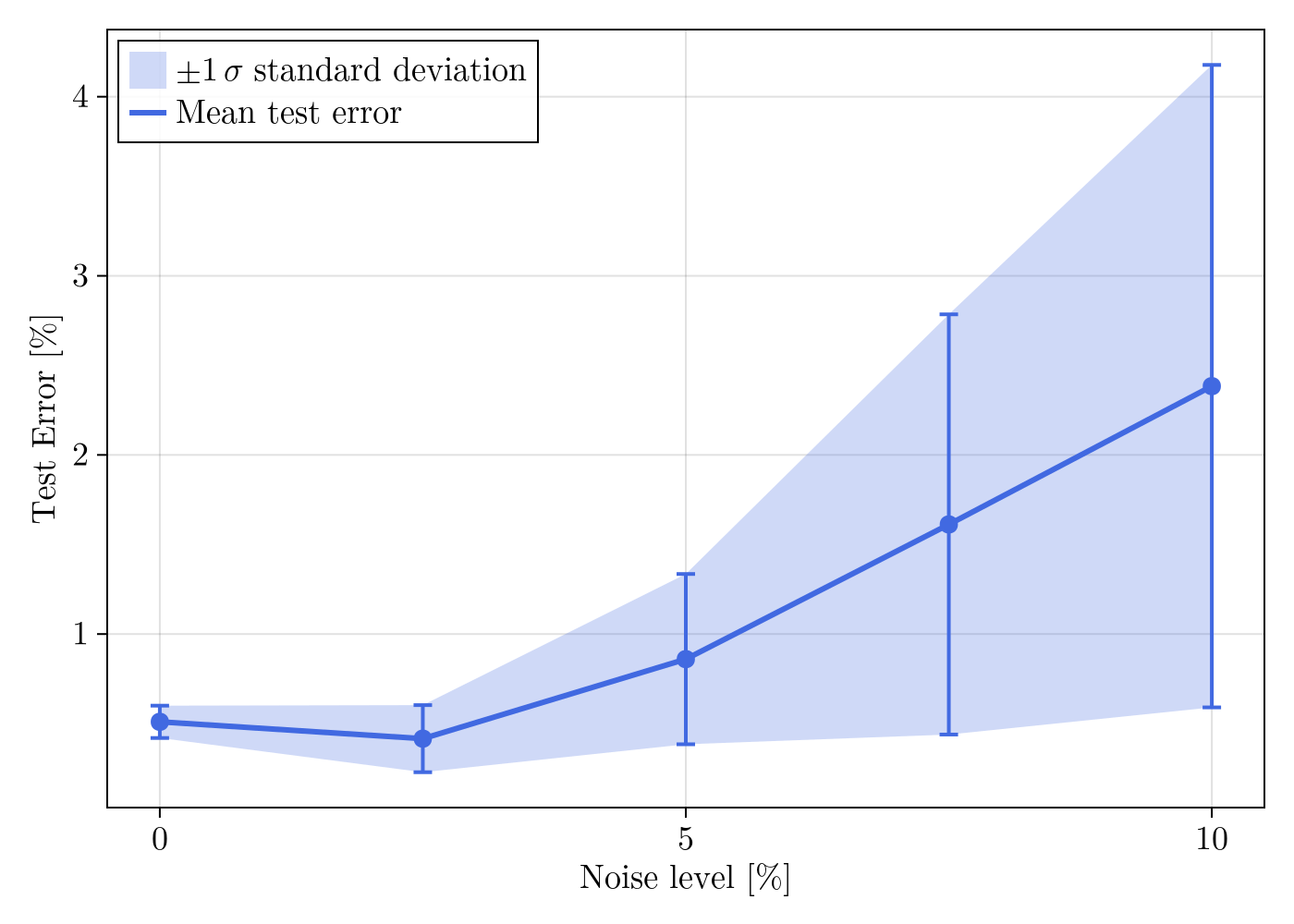}
\caption{Root mean square error of learned neural networks with increasing noise level.}
\label{fig:D1_Noise}
\end{figure}

\begin{figure}
\centering
\includegraphics[width=0.5\textwidth]{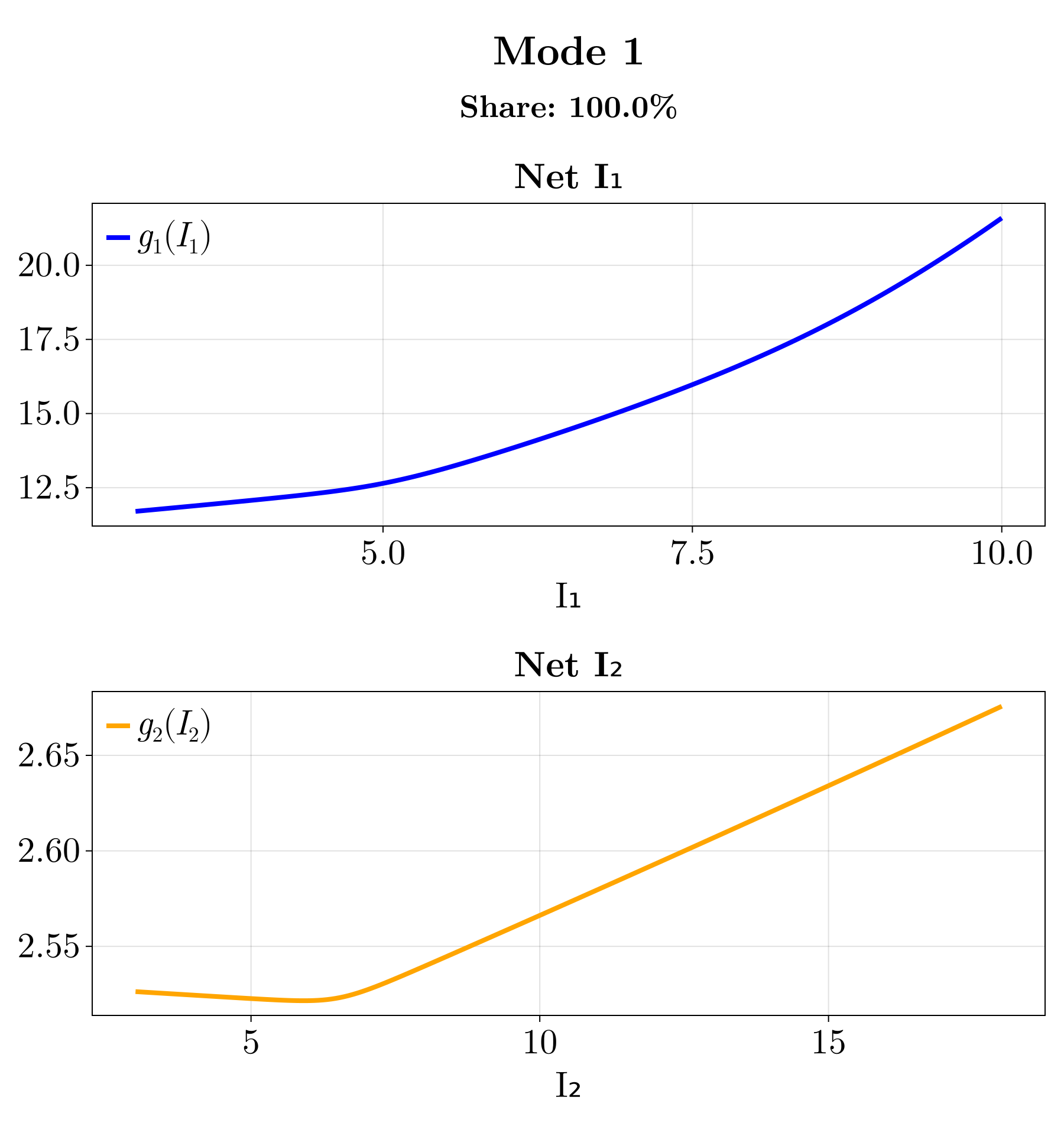}
\caption{An alternative structure of a neural network for the polynomial material model.}
\label{fig:D1_Noise_Architecture}
\end{figure}

\begin{figure}
\centering
\includegraphics[width=0.8\textwidth]{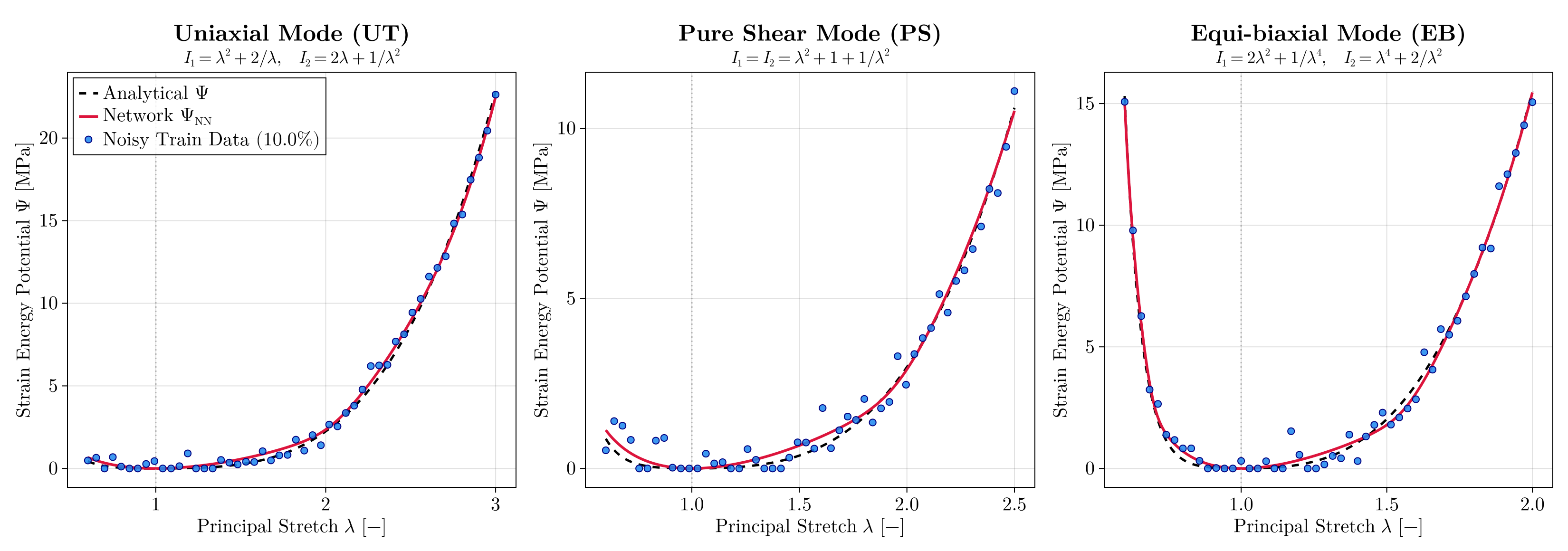}
\caption{Comparison of the training data with noise and the neural network predictions.}
\label{fig:D1_Noise_Incompressible_Deformation_Modes}
\end{figure}

The test root mean square error with increasing noise level is shown in Figure~\ref{fig:D1_Noise}. The mean and standard deviation of the error across five runs are shown. The error is computed as the mean squared error between the predicted and true strain energy values. Noise levels below 5\% have only a minor effect on the identification of the reference material model. This robustness may reflect the network's ability to recover the underlying structure of the strain energy despite the noise.
However, for a noise level of 5\% and above, the network sometimes identifies a different material model, consisting of only one mode. An identified material model for a noise level of 10\% is shown in Figure~\ref{fig:D1_Noise_Architecture}.
This model consists of a single mode that couples both invariants.
The learned strain energy function is compared to the true strain energy function along the loading conditions in Figure~\ref{fig:D1_Noise_Incompressible_Deformation_Modes}. The model is able to represent the training data reasonably well, but fails to capture the underlying structure of the strain energy function. 
As a result, a higher test error of 1.4\% is reported.
Inspecting the learned components can reveal unexpected behavior, such as the kink in the $I_2$ dependence at approximately $ I_2=6.0 $. Such observations can guide additional experiments to improve model accuracy and generalization. This diagnostic information is less directly accessible in a black-box constitutive neural network.


\subsection{Mooney-Rivlin material model}
\label{sec:MooneyRivlin}
As a second supervised benchmark, we consider a compressible Mooney-Rivlin material with the strain energy function
\begin{equation}
\Psi = C_{10} (I_1 J^{-2/3} - 3) + C_{01} (I_2 J^{-4/3} - 3) + D (J-1)^2
\end{equation}
with $C_{10} = \SI{0.5}{MPa}$, $C_{01} = \SI{0.2}{MPa}$, and $D = \SI{5.0}{MPa}$. The first two terms describe the isochoric response, whereas the last term describes the volumetric response.

The isochoric split removes the change in volume from the first two invariants.
For a deformation gradient in three dimensions, the factors $J^{-2/3}$ and $J^{-4/3}$ compensate for the volumetric scaling of $I_1$ and $I_2$, respectively. Consequently, $I_1 J^{-2/3}$ and $I_2 J^{-4/3}$ describe shape changes independently of the change in volume, while $J$ controls the volumetric part of the response. 
This split provides a useful test for the proposed architecture: the network must learn both the non-integer powers of $J$ that couple the isochoric invariants to the original coordinates and the separate volumetric contribution.

The material model is not convex in $I_1$, $I_2$, and $J$. This benchmark therefore also tests the network's ability to approximate a nonconvex reference energy.

For the training data, we sample the admissible deformation domain with
$I_1 \in [3.0,20.0]$, $I_2 \in [3.0,20.0]$, and $J \in [0.25,4.0]$.
Half of the samples have $J \in [0.25,1.0]$ and half have $J \in [1.0,4.0]$, so that both compression and extension are represented. The dataset contains 500 training points in total.

The training details are summarized in \tabref{tab:D3_TrainingDetails}. We choose a large value of $\lambda_\textrm{3 cross}$ to discourage three-way invariant couplings, which are absent from the reference model. The network is initialized with 5 modes, each consisting of 3 subnetworks for the invariants $I_1$, $I_2$, and $J$.  The last mode is a pure volumetric mode with $f_{m, 1} = f_{m, 2} = 0$ and $f_{m, 3} \neq 0$. Nevertheless, the network can prune this mode if it is not needed.
Each subnetwork consists of 20 hidden neurons, resulting in a total of 806 trainable parameters.

\begin{table}[htbp]
\centering
\caption{Training details for the identification of the Mooney-Rivlin material model.}
\label{tab:D3_TrainingDetails}
\begin{tabular}{ll}
\toprule
Learning rate & $5 \times 10^{-2}$ \\
Optimizer & Adam until 37500 epochs \\
& L-BFGS for the last 2500 epochs \\
Total number of epochs & 40000 \\
Regularization & $\lambda_\textrm{L1} =  1 \times 10^{-3}$ \\
& $\lambda_\textrm{3 cross} = 10.0$ \\
& \rev{linear} ramp up until 10\,000 epochs \\
Mode pruning & After 10\,000 epochs \\
\bottomrule
\end{tabular}
\end{table}

\begin{figure}
\centering
\includegraphics[width=0.8\textwidth]{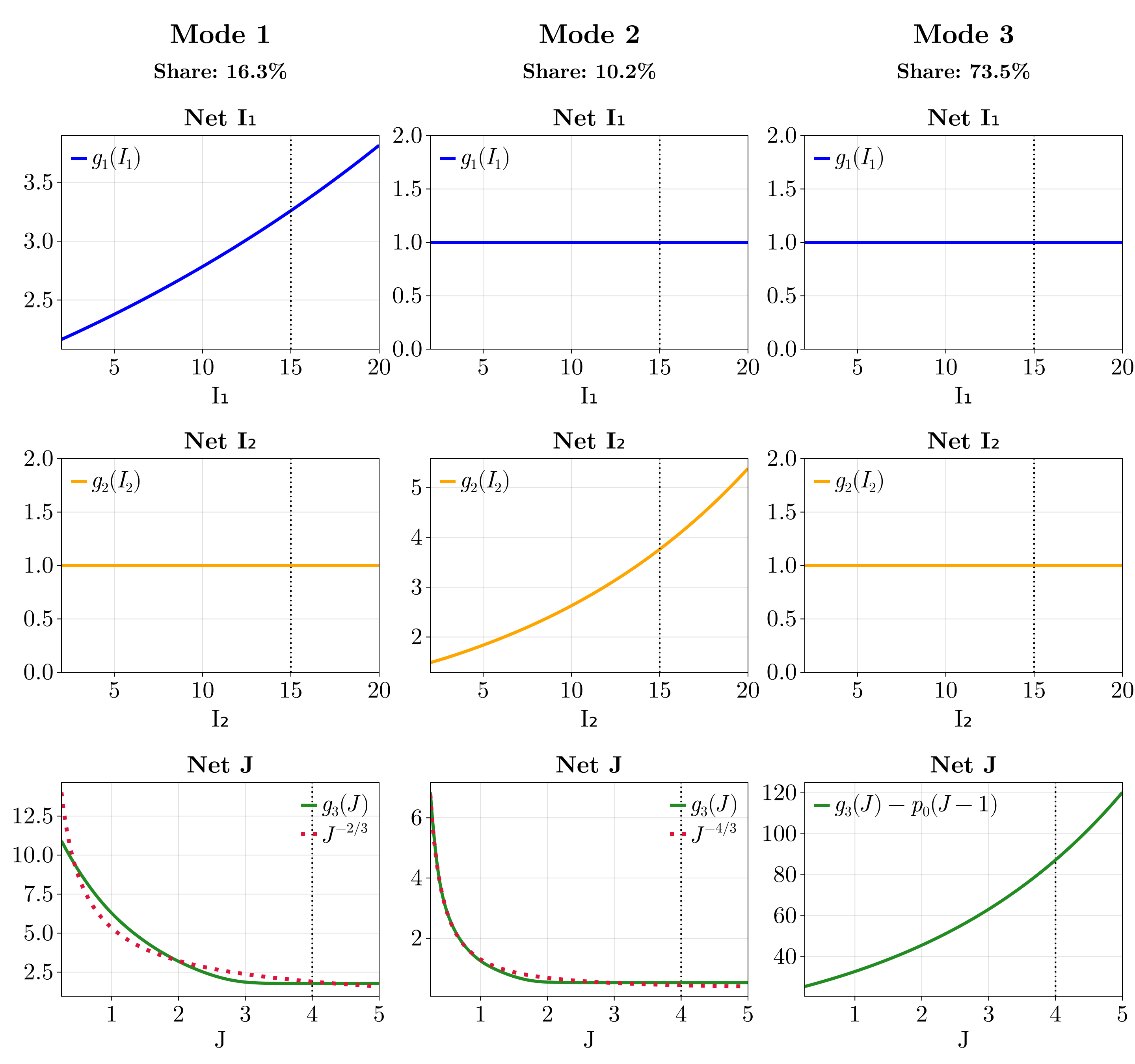}
\caption{Identified material model for the Mooney-Rivlin material model with isochoric split.}
\label{D3_Architecture}
\end{figure}

The identified network is presented in Figure~\ref{D3_Architecture}. The network consists of 3 modes with the same structure.
For the networks representing the Jacobian, the analytical expressions are presented as dotted red lines in the figure. 
The powers $J^{-2/3}$ and $J^{-4/3}$ are learned with reasonable accuracy, demonstrating the ability of the network to learn non-integer powers of the invariants. The error on the training data is 2.6\%, which is higher than in the previous example. The higher error may reflect the mismatch between the convex architecture and the reference energy, which is not convex in $I_1$, $I_2$, and $J$.

In library-based approaches such as EUCLID, the invariant powers must be included in the candidate library before training. Recovering these terms therefore requires a sufficiently informative library.
The quadratic volumetric term $(J-1)^2$ is shared across the individual modes. The left branch for $J < 1$ is included in the first two modes. This is reasonable, as we did not force the network to learn a symmetric response for compression and extension. The inclusion of this compression branch may also explain the slight differences in the shapes of the learned functions of $J$. Mathematically, the separation of the energy is not unique.

In Figure~\ref{D3_Architecture}, the range of the invariants is increased beyond the training domain to showcase the reasonable extrapolation behavior of the network. The end of the training domain is marked with a black dashed line.

\begin{figure}
\centering
\includegraphics[width=1.0\textwidth]{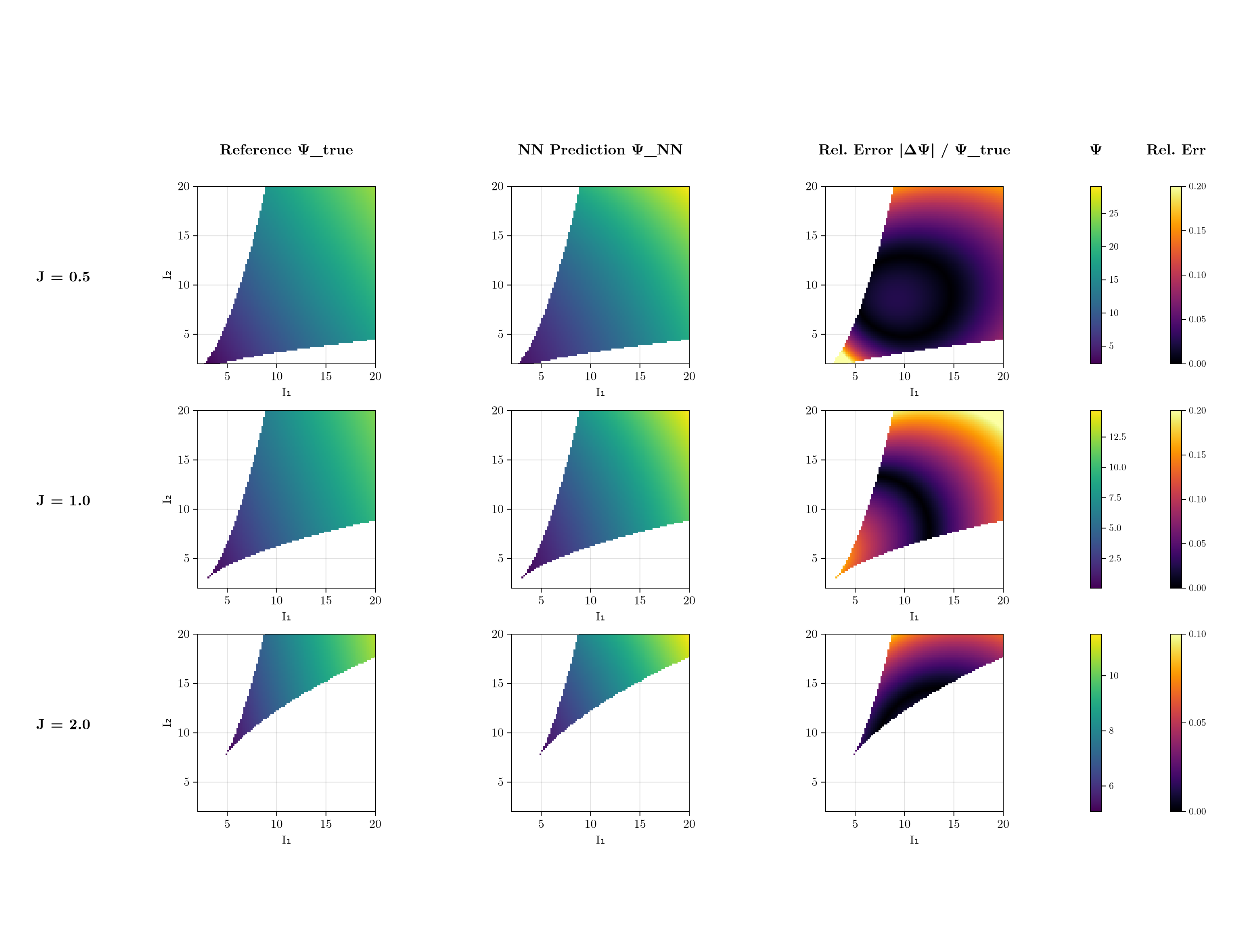}
\caption{Comparison of the learned strain energy function with the true strain energy function of the Mooney-Rivlin material model for different values of $J$.}
\label{D3_Energy_Landscape}
\end{figure}

In Figure~\ref{D3_Energy_Landscape}, the learned strain energy function is visualized in the invariant plane. 
The learned energy agrees reasonably well with the reference energy within the training domain. Errors increase outside this domain, particularly at large $I_1$ and $I_2$.

\subsection{Unsupervised identification}
\label{sec:Unsupervised}
In the unsupervised benchmark, we identify the constitutive model from full-field deformation data and global reaction forces, quantities accessible in experiments. Training uses equilibrium and reaction-force residuals without local stress or strain-energy labels.

We use a Neo-Hookean reference material with the strain energy
\begin{equation}
\Psi = \frac{\mu}{2} (I_1 - 3 - 2 \ln(J)) + \frac{\lambda}{2} (J-1)^2
\end{equation}
with the material parameters $\mu = \SI{4}{MPa}$ and $\lambda = \SI{6}{MPa}$.

To assess robustness to imperfect measurements, additive noise with a level
of $2\%$ is applied to the distances of the invariant fields from the
reference state, i.e., to $(I_1-3,I_2-3,J-1)$. The resulting noisy kinematic
fields are used as the only local input to the identification procedure; the
reaction forces remain the global measurement used to constrain the learned
model. Training starts with Adam, during which the regularization weights are
ramped up and redundant modes are pruned. The resulting sparse model is then
fine-tuned with L-BFGS.

\begin{table}[htbp]
\centering
\caption{Training details for the identification of the Neo-Hookean material model.}
\label{tab:Unsupervised_TrainingDetails}
\begin{tabular}{ll}
\toprule
Learning rate & $5 \times 10^{-2}$ \\
Optimizer & Adam until 1000 epochs \\
& L-BFGS until 3000 epochs \\
Total number of epochs & 3000 \\
Regularization & $\lambda_\textrm{L1} =  1 \times 10^{-1}$ \\
& $\lambda_\textrm{3 cross} = 1.0$ \\
& \rev{linear} ramp up until 500 epochs \\
Mode pruning & After 500 epochs \\
\bottomrule
\end{tabular}
\end{table}

\begin{figure}
\centering
%% Creator: Inkscape 1.4.4 (dcaf3e7, 2026-05-05), www.inkscape.org
%% Accompanies image file 'Geometry.pdf' (pdf, eps, ps)
\begingroup%
  \makeatletter%
  \providecommand\color[2][]{%
    \errmessage{(Inkscape) Color is used for the text in Inkscape, but the package 'color.sty' is not loaded}%
    \renewcommand\color[2][]{}%
  }%
  \providecommand\transparent[1]{%
    \errmessage{(Inkscape) Transparency is used (non-zero) for the text in Inkscape, but the package 'transparent.sty' is not loaded}%
    \renewcommand\transparent[1]{}%
  }%
  \providecommand\rotatebox[2]{#2}%
  \newcommand*\fsize{\dimexpr\f@size pt\relax}%
  \newcommand*\lineheight[1]{\fontsize{\fsize}{#1\fsize}\selectfont}%
  \ifx\svgwidth\undefined%
    \setlength{\unitlength}{366.01462537bp}%
    \ifx\svgscale\undefined%
      \relax%
    \else%
      \setlength{\unitlength}{\unitlength * \real{\svgscale}}%
    \fi%
  \else%
    \setlength{\unitlength}{\svgwidth}%
  \fi%
  \global\let\svgwidth\undefined%
  \global\let\svgscale\undefined%
  \makeatother%
  \begin{picture}(1,0.39676545)%
    \lineheight{1}%
    \setlength\tabcolsep{0pt}%
    \put(0,0){\includegraphics[width=\unitlength,page=1]{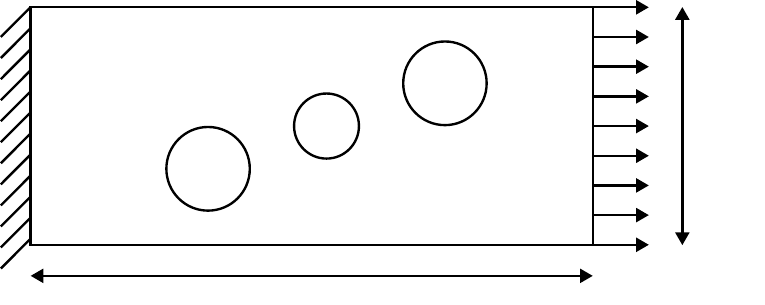}}%
    \put(0.33973107,0.00192116){\color[rgb]{0,0,0}\makebox(0,0)[lt]{\lineheight{1.25}\smash{\begin{tabular}[t]{l}100 mm\end{tabular}}}}%
    \put(0.91283325,0.22029666){\color[rgb]{0,0,0}\makebox(0,0)[lt]{\lineheight{1.25}\smash{\begin{tabular}[t]{l}40 mm\end{tabular}}}}%
  \end{picture}%
\endgroup%

\caption{Plate geometry for the unsupervised identification benchmark.}
\label{EUCLID_Geometry}
\end{figure}

As training data, we use the kinematic deformation fields and the global reaction forces from a finite-element simulation of a thin plate with three holes under uniaxial tension. The geometry of the plate is shown in Figure~\ref{EUCLID_Geometry}. \revv{The position and size of the holes is presented in Table~\ref{tab:hole_geometry}.} \rev{The height is \SI{2}{mm}. On the left hand side a zero Dirichlet boundary condition is applied whereas on the right hand side non-zero Dirichlet boundary condition is prescribed. The reaction force is measured at the right-hand side of the plate. A finite element mesh with 1357 tetrahedral elements is used.} The plate is stretched in the horizontal direction. The training data consist of snapshots at 25, 50, 75, and 100\% elongation of the plate.

\begin{table}[htbp]
\centering
  \caption{Geometry of the three circular holes in the plate.}
\label{tab:hole_geometry}
\begin{tabular}{lccc}
\toprule
Hole & Midpoint $x$ [mm] & Midpoint $y$ [mm] & Radius $r$ [mm] \\
\midrule
1 & 30.0 & 12.8 & 7.2 \\
2 & 70.0 & 27.2 & 7.2 \\
3 & 50.0 & 20.0 & 5.6 \\
\bottomrule
\end{tabular}
\end{table}

\begin{figure}
\centering
\includegraphics[width=0.5\textwidth]{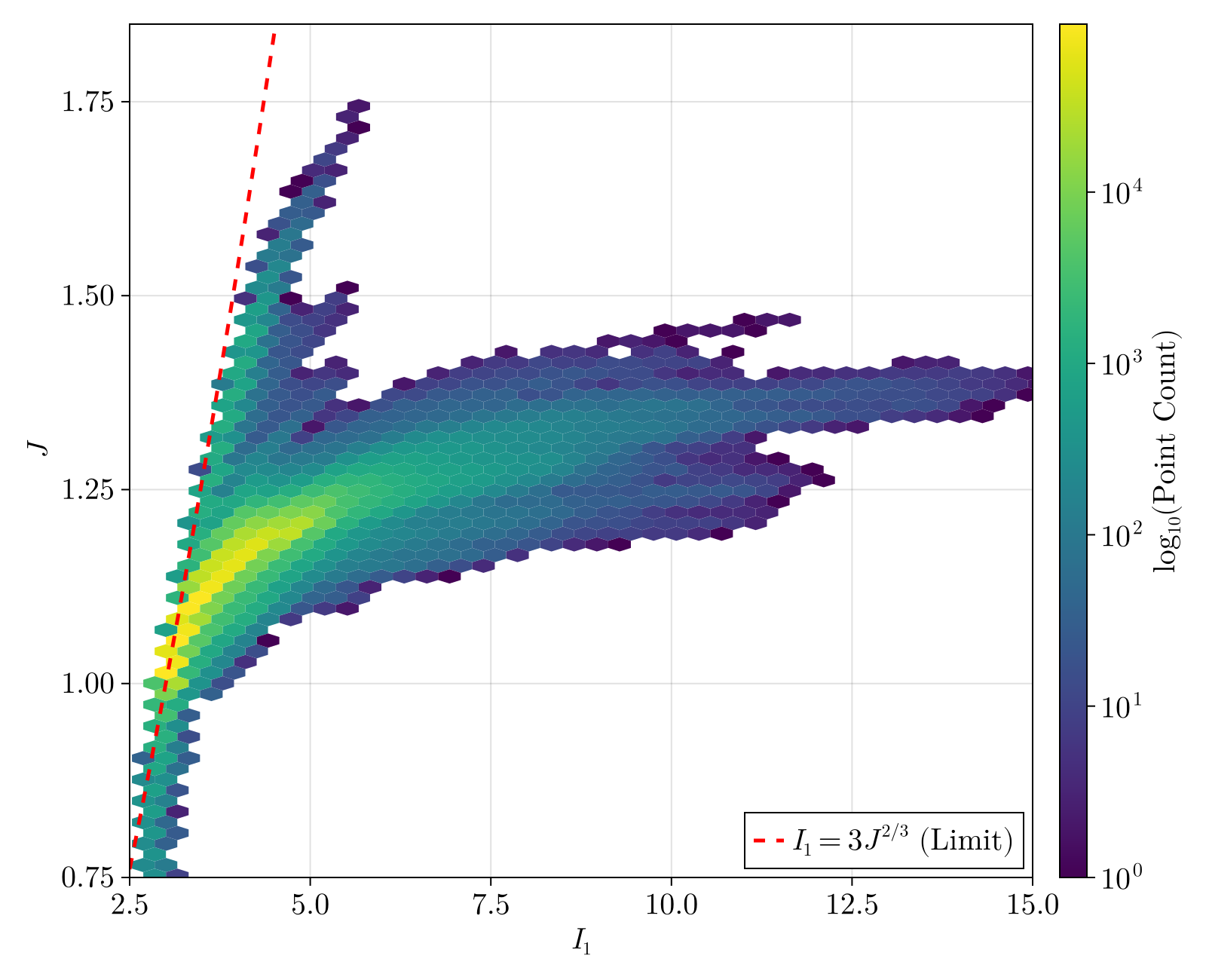}
\caption{Deformation states in the training data.}
\label{EUCLID_DeformationStates}
\end{figure}

In Figure~\ref{EUCLID_DeformationStates}, the deformation states in the training data are visualized. The color indicates the number of data points in the training dataset for a given combination of invariants $I_1$ and $J$. \rev{No data points were found in the white region.} The training data is concentrated in the region defined by the invariants $I_1 \in [3, 5]$ and $J \in [1.0, 1.25]$.

\begin{figure}
\centering
\includegraphics[width=0.5\textwidth]{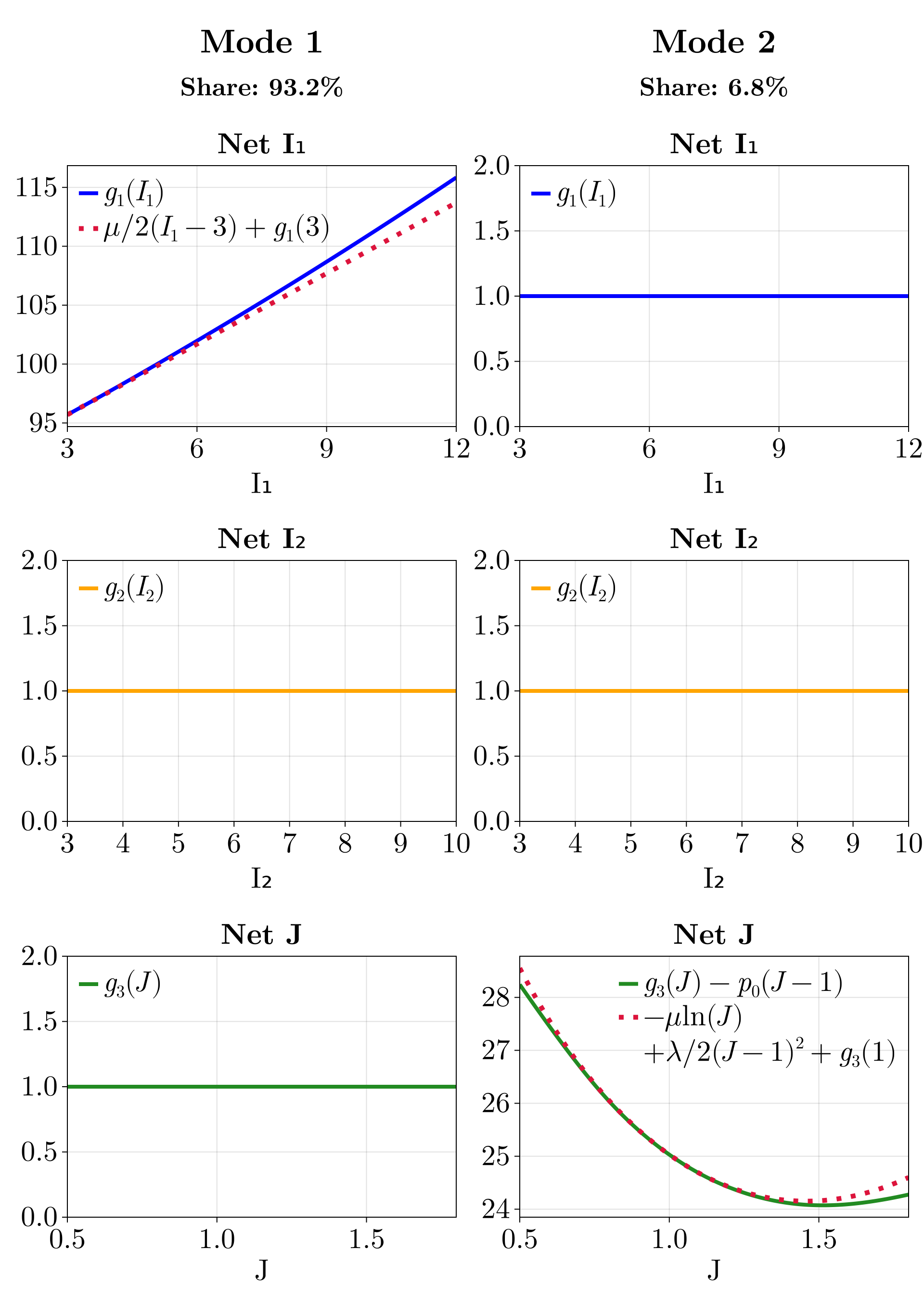}
\caption{Identified constitutive model for the unsupervised benchmark.}
\label{EUCLID_Model_Architecture}
\end{figure}

The training details are summarized in Table~\ref{tab:Unsupervised_TrainingDetails}. The network is initialized with 5 modes, each consisting of 3 subnetworks for the invariants $I_1$, $I_2$, and $J$. 
The identified constitutive model after training of the neural network is shown in Figure~\ref{EUCLID_Model_Architecture}. The model can approximate the data reasonably well with two active modes. The analytical functions for the active terms are shown as dotted red lines in the figure. The network recovers the structure of the reference material model, correctly identifying the absence of any contribution from $I_2$ or coupling between invariants.

\begin{figure}
\centering
\includegraphics[width=0.5\textwidth]{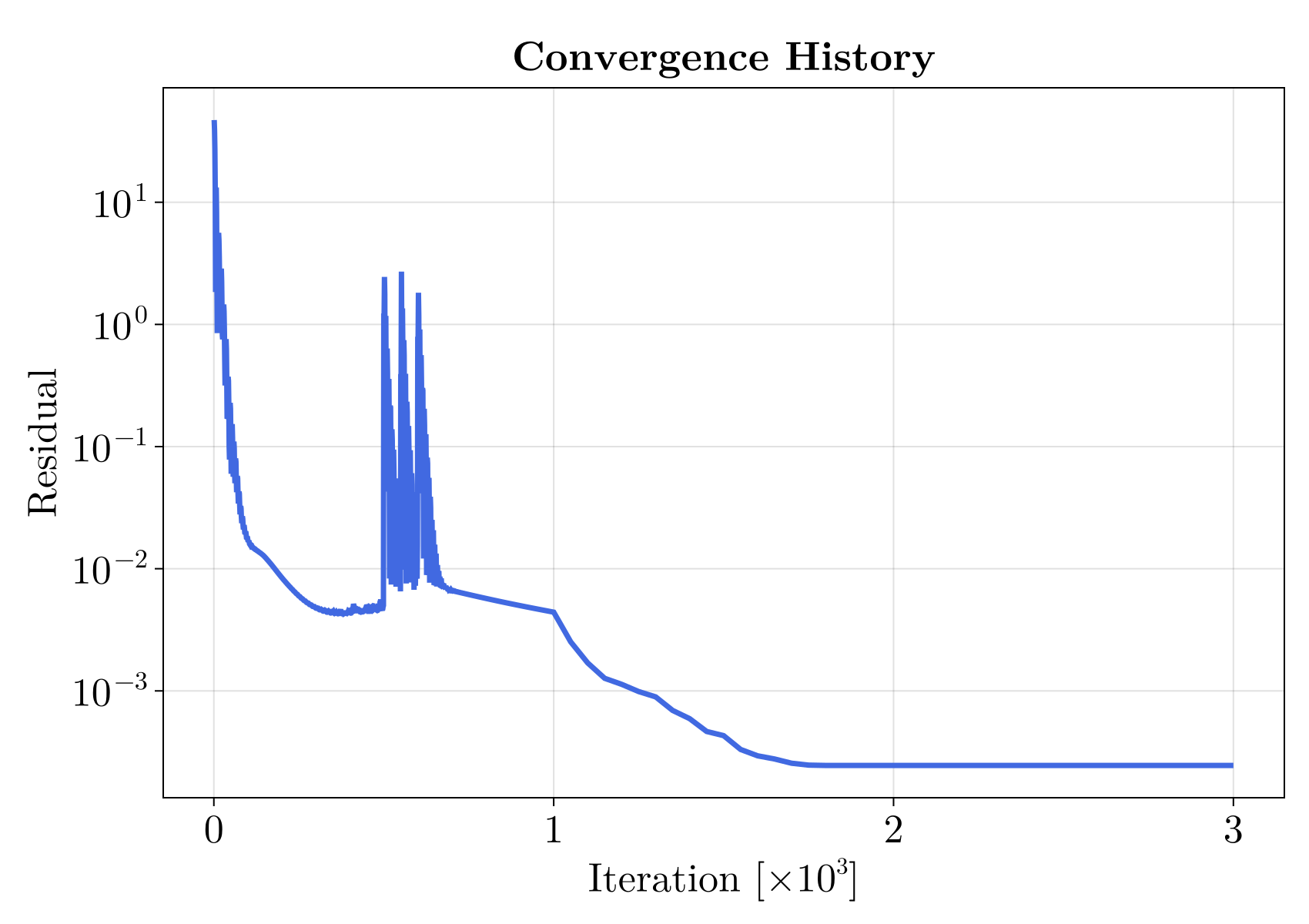}
\caption{Convergence history of the unsupervised learning process.}
\label{EUCLID_Convergence_History}
\end{figure}

The convergence history is shown in Figure~\ref{EUCLID_Convergence_History}. \rev{The residual is given in Equation~\eqref{eq:EUCLIDLoss}.}
Adam first identifies the overall structure of the model, including the active invariant dependencies, while the subsequent L-BFGS stage, beginning after 1000 epochs, substantially reduces the remaining force and equilibrium residuals.

\begin{figure}
\centering
\includegraphics[width=0.5\textwidth]{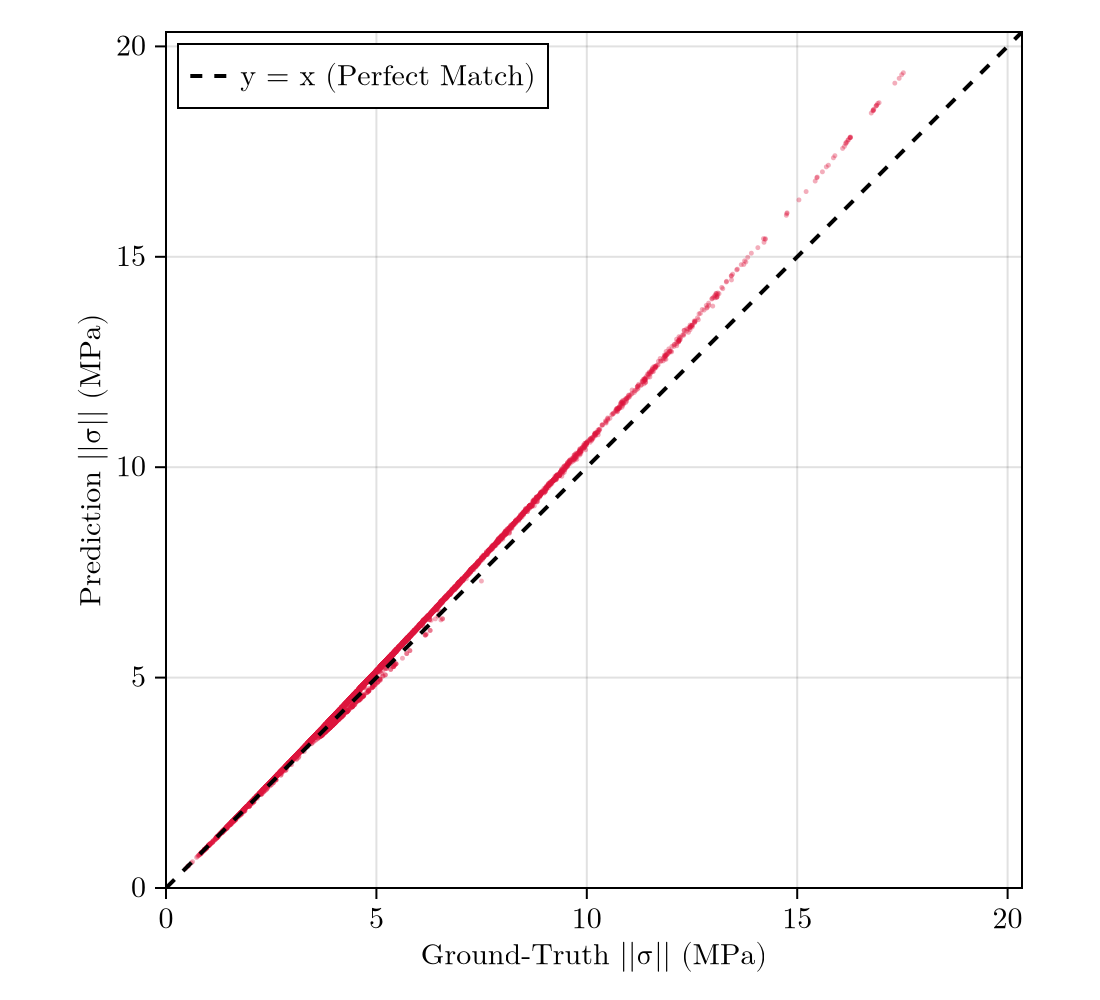}
\caption{Comparison of predicted and reference stress distributions at 50\% elongation.}
\label{EUCLID_Training_Stress_Correlation}
\end{figure}

The predicted and reference stress distributions at 50\% elongation are compared in Figure~\ref{EUCLID_Training_Stress_Correlation}. The model reproduces the overall stress response well, with $R^2=0.977$. The model overpredicts the stress especially for high stress values.

\begin{figure}
\centering
\includegraphics[width=1.0\textwidth]{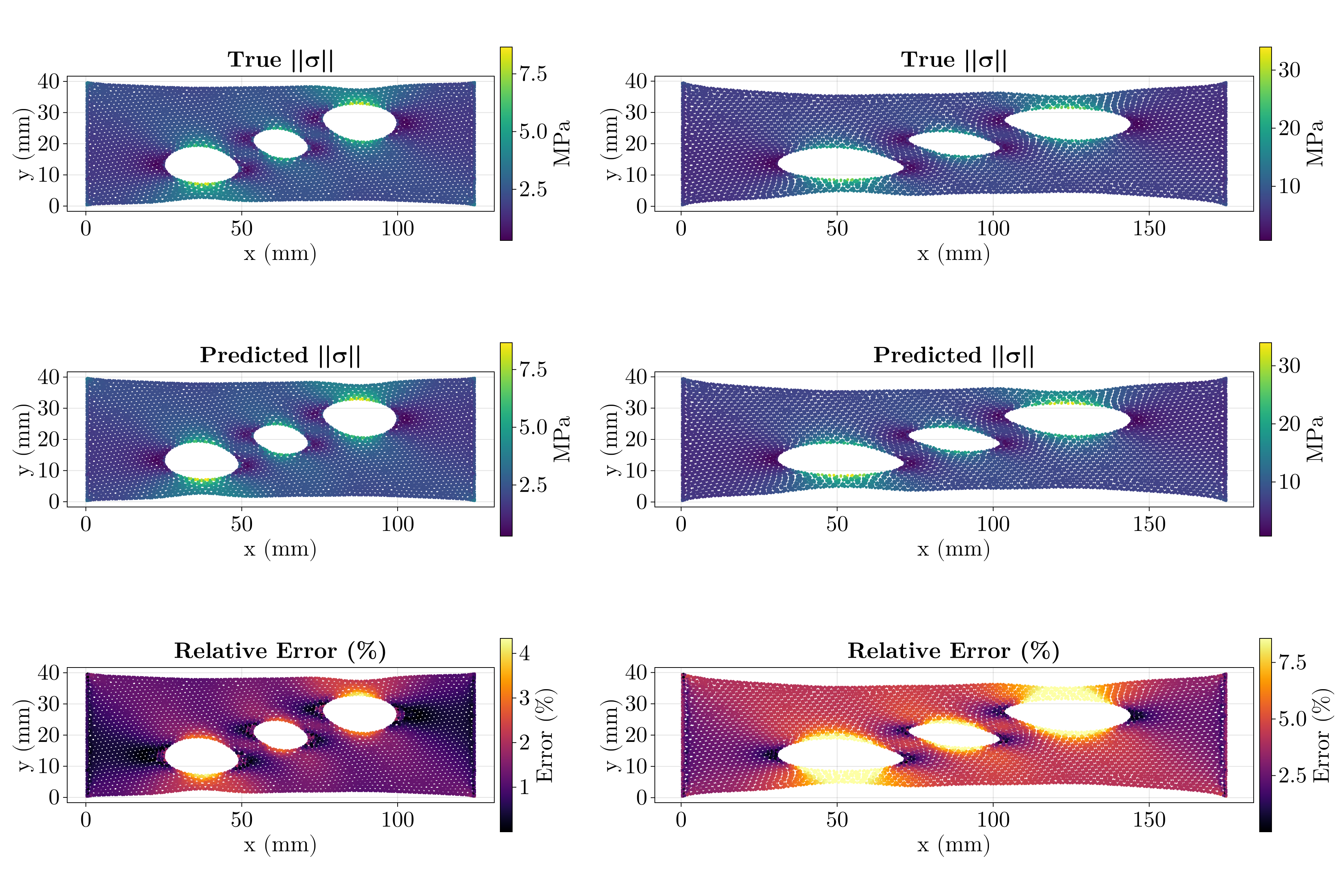}
\caption{Stress field maps at 25\% and 75\% elongation.}
\label{EUCLID_Training_Stress_Field_Maps}
\end{figure}

Figure~\ref{EUCLID_Training_Stress_Field_Maps} compares the predicted and reference stress fields at 25\% and 75\% elongation. The spatial stress patterns agree well at both load levels, including the localization around the holes. The largest discrepancies occur in regions with high stress. These deviations are consistent with the limited number of training states and the concentration of the measured deformation data in a relatively narrow region of invariant space. \rev{A reweighting of the training data with respect to the data density might reduce these deviations even further.}

\section{Discussion}
\label{sec:Discussion}
This work presents a constitutive neural-network architecture that combines the
approximation capability of neural networks with the structural interpretability
of sparse constitutive models, such as those identified by EUCLID. The strain energy is represented as a sum of
modes, each formed from one-dimensional input-convex subnetworks in a single invariant. Sparsity is
promoted by regularization and mode pruning. This combination provides
an interpretable decomposition without requiring the relevant functional forms
to be specified in advance through a fixed library.

\rev{The proposed architecture can be seen as a complementary
approach to Physics-augmented Neural Networks (PANNs), Constitutive Neural Networks (CANNs), and Kolmogorov-Arnold networks (KANs). PANNs and CANNs
can provide a highly expressive approximation while enforcing selected physical
properties, but their internal parameters do not necessarily reveal the
invariant dependencies learned from the data. KANs also make one-dimensional functions accessible
for visualization, although their multilayer compositions can make the
relationship between these functions and the strain energy difficult to
interpret. In contrast, the present approach learns one-dimensional functions
without prescribing their shapes, while the sparse sum-of-products structure
makes the retained invariant couplings explicit. It therefore aims to preserve
much of the expressivity of neural networks while providing a decomposition
that can be inspected and related to established constitutive models. This
interpretability should not be understood as uniqueness: different
decompositions may represent the same constitutive response, and the selected
terms should consequently be interpreted as a compact representation supported
by the data rather than as a unique identification of physical mechanisms.}

The supervised benchmarks demonstrate the ability of the proposed approach to
recover both the structure and the functional behavior of representative
hyperelastic energies. The architecture also approximates the isochoric split of the Mooney-Rivlin model, including invariant powers that must be specified in advance in library-based approaches.
These results illustrate the main advantage over a prescribed library: the network can learn the shape of the
one-dimensional functions while the regularization determines which invariant
couplings are retained. At the same time, the observed extrapolation errors
outside the training domain show that the physical constraints and convexity of
the architecture do not eliminate the need for sufficiently informative
training data.

The unsupervised experiment further shows that the same representation can be
identified from full-field kinematics and global reaction forces without local
stress labels. The learned model reproduces the measured force response and
captures the principal stress-field patterns, although deviations remain in
regions with large stresses and sparse deformation data. 

Several limitations motivate future work. The present study considers isotropic
materials described by three invariants. The framework can in
principle be extended to larger sets of invariants, including anisotropic or
generalized invariants, but this will increase both the number of possible
couplings and the amount of deformation data required for reliable model
selection. \rev{A suitable sparsity strategy would have to be developed.}

\rev{An important next step is the application to experimental data, which
introduce noise, calibration errors, incomplete fields, and uncertainty in the
applied boundary conditions; robust training, uncertainty quantification, and
validation on independent loading paths will therefore be essential for
assessing the generalizability and identifiability of the sparse decomposition.
The architecture could also be extended to families of materials by
introducing a material embedding that learns a
low-dimensional representation from multiple experiments \cite{mabileau_learning_2027}. Such an extension
could enable information sharing and interpolation between material states and could connect sparse constitutive discovery with surrogate
modeling for material design and multi-material finite-element simulations.}

\section*{Acknowledgment}
Funded by the European Union (ERC, Gen-TSM, 101124463). Views and opinions expressed are however those of the author(s) only and do not necessarily reflect those of the European Union or the European Research Council Executive Agency. Neither the European Union nor the granting authority can be held responsible for them.

\section*{Data availability}
The code is available at \url{https://github.com/ikm-luh/Sparse_PGD_NN_SourceCode}.

\section*{Declaration of gnerative AI and AI-assisted technologies}
During the preparation of this work the author(s) used Gemini 3.8 Flash and OpenAI GPT 6.1 Sol in order to improve the readability and language of the manuscript and assist with implementing and debugging the proposed algorithm. After using this tool/service, the author(s) reviewed, tested, and edited the content and code as needed and take(s) full responsibility for the content of the published article.
\appendix

\addcontentsline{toc}{chapter}{Bibliography}
\printbibliography
%\bibliographystyle{plain}
%\bibliography{bib_short}

\end{document}